\documentclass[aps,rmp,reprint,amsmath,amssymb,graphicx,longbibliography]{revtex4-1}
\usepackage{float}
\usepackage{amssymb,amsmath,graphicx,xcolor,hyperref}
\usepackage[normalem]{ulem}	
\usepackage{parskip}

\begin{document}

\title{Hybrid Halide Perovskites: Fundamental Theory and Materials Design}

\author{Marina R. Filip$^1$}
\affiliation{$^1$Department of Materials, University of Oxford, Parks Road, Oxford OX1 3PH, United 
Kingdom}
\author{George Volonakis$^1$}
\affiliation{$^1$Department of Materials, University of Oxford, Parks Road, Oxford OX1 3PH, United 
Kingdom}
\author{Feliciano Giustino$^{1,2}$}
\email{feliciano.giustino@materials.ox.ac.uk}
\affiliation{$^1$Department of Materials, University of Oxford, Parks Road, Oxford OX1 3PH, United 
Kingdom}
\affiliation{$^2$Department of Materials Science and Engineering, Cornell University, Ithaca, New York 
14853, USA}

\newcommand{\hbindex}[1]{\hl{#1}\index{#1}}  

\begin{abstract}
Hybrid organic-inorganic halide perovskites have emerged as a disruptive new class of materials, 
exhibiting optimum properties for a broad range of optoelectronic applications, most notably for 
photovoltaics. The first report of highly efficient organic-inorganic perovskite solar cells in 2012 
\citep{Lee2012} marked a new era for photovoltaics research, reporting a power 
conversion efficiency of over 10\%~\citep{NREL}. Only five years after this discovery, perovskite 
photovoltaic devices have 
reached a certified efficiency of 22.7\%, making them the first solution processable technology 
to surpass thin film and multi-crystalline silicon solar cells~\citep{NREL}. The remarkable 
development of perovskite solar cells is due to the ideal optoelectronic properties of 
organic-inorganic lead-halide perovskites. The prototypical compound, methylammonium lead  iodide, 
CH$_3$NH$_3$PbI$_3$~\citep{Stranks2015} is a direct band gap semiconductor with a band gap in the 
visible, high charge carrier mobility, long diffusion length and low excitonic binding 
energy~\citep{Johnston2016}. Due to these ideal properties, CH$_3$NH$_3$PbI$_3$ is also drawing 
interest across many other applications beyond photovoltaics, such as light emitting 
devices~\citep{Tan2014}, lasers~\citep{Wehrenfennig2014}, photocatalysts~\citep{Chen2015} and  
transistors~\citep{Ward2017}. 
The continued progress of metal-halide perovskite optoelectronics relies not only on a detailed
understanding of the electronic and optical properties of materials in this class, but also on the
development of practical strategies to tune their properties by controlling parameters such as
chemical composition. In this context, {\it ab initio} computational modelling can play a key role
in providing a physical interpretation of experimental measurements, and guiding the design of novel
halide perovskites with tailored properties.
In this chapter we will present an account of the contributions to this fast developing field of
research from our computational modelling group. The chapter is organized in two sections. The first
section focuses on the structural and optoelectronic properties of CH$_3$NH$_3$PbI$_3$. Here, we
expand on some of the challenging aspects of modelling the electronic and vibrational properties of
CH$_3$NH$_3$PbI$_3$, and discuss the main theoretical results alongside experimental data. The
second section discusses the recent computationally-led materials design of novel halide
perovskites, and the principal challenges in replacing Pb$^{2+}$ in CH$_3$NH$_3$PbI$_3$ by non-toxic
elements.
\end{abstract}

\maketitle

\begin{section}{Methylammonium Lead Iodide}

\begin{figure*}[ht!]
\begin{center}
\includegraphics[width=0.9\textwidth]{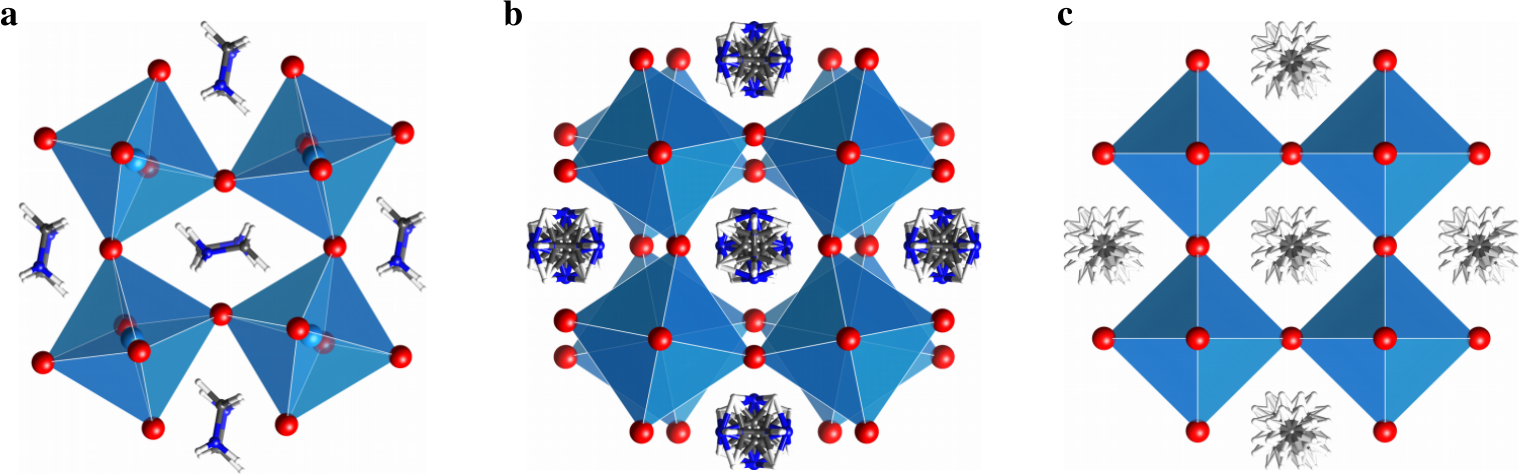}
\caption{{\bf Temperature dependent crystal structure of CH$_3$NH$_3$PbI$_3$.} Polyhedral
representation of the low temperature orthorhombic $Pnma$ structure ({\bf a}), the tetragonal
$I4/mcm$ structure ({\bf b}) and the cubic $Pm\bar{3}m$ structure ({\bf c}). In all three panels
the Pb atoms are represented by the light blue spheres at the center of the octahedra, the I atoms
are the red spheres at the corners of the octahedra, and the C, N and H atoms are represented by the
dark grey, dark blue, and white spheres, respectively. The figures were constructed using the
lattice parameters and atomic positions reported by~\cite{Weller2015}.\label{fig1}}
\end{center}
\end{figure*}

Methylammonium lead-iodide, CH$_3$NH$_3$PbI$_3$, belongs to the ABX$_3$ perovskite structural
family~\citep{Poglitsch1987}. As shown in FIG.~\ref{fig1}, the Pb$^{2+}$ and I$^-$ ions form a
three-dimensional network of corner-sharing octahedra. The organic CH$_3$NH$_3^+$ cations occupy the
center of the cuboctahedral cavities enclosed by the inorganic PbI$_6$ network
\citep{Poglitsch1987}. The crystal structure of CH$_3$NH$_3$PbI$_3$ is strongly dependent on
temperature and undergoes two phase transitions~\citep{Poglitsch1987, Baikie2013,Stoumpos2013,
Weller2015}: from the low temperature orthorhombic $Pnma$ structure to the tetragonal $I4/mcm$
structure at 160~K, and the cubic $Pm\bar{3}m$ structure at 315-330~K. As shown in FIG.~\ref{fig1},
all phases of CH$_3$NH$_3$PbI$_3$ maintain the corner sharing connectivity of the PbI$_6$ octahedra,
however the degree of octahedral tilting is reduced as the temperature increases. In addition, the
organic CH$_3$NH$_3^+$ cations exhibit ordered  orientations throughout the unit cell only in the
orthorhombic phase, while in the tetragonal and cubic phase the cations are orientationally
disordered~\citep{Poglitsch1987, Baikie2013, Stoumpos2013, Weller2015}.

The excited state properties of CH$_3$NH$_3$PbI$_3$ have been intensely investigated over the last
five years. The optical absorption spectrum of CH$_3$NH$_3$PbI$_3$ is characteristic of a direct
band gap semiconductor~\citep{Herz2016}, with a sharp absorption onset around 1.6~eV. At low
temperature, the optical absorption spectrum exhibits an excitonic peak, with an exciton binding
energy of $\sim$20~meV. This feature becomes less visible as the temperature increases
\citep{DInnocenzo2014,Myiata2015,Herz2016}. In addition, the absorption lineshape at room
temperature exhibits an Urbach tail of 15~meV~\citep{deWolf2014}, unexpectedly low given the
disordered character of the CH$_3$NH$_3^+$ cation. The optical band gap of CH$_3$NH$_3$PbI$_3$
increases with temperature from 1.62~eV at 4~K to 1.67 eV at 160~K. At the phase transition
temperature between the orthorhombic and tetragonal phase, 160~K, the band gap decreases sharply to
approximately 1.58~eV, and then continues to increase beyond this temperature, as in the
orthorhombic phase~\citep{Millot2015}.

Free electrons and holes dominate over bound exciton pairs upon photoexcitation, an ideal
characteristic for application in solar energy conversion~\citep{DInnocenzo2014,Herz2016}. The
photogenerated charge carriers are long lived in CH$_3$NH$_3$PbI$_3$ and exhibit charge carrier
mobilities of approximately 35~cm$^2(Vs)^{-1}$ \citep{Johnston2016} at room temperature. Bimolecular
recombination in CH$_3$NH$_3$PbI$_3$ is found to be the predominant mechanism for charge carrier
recombination at standard photovoltaic device operation conditions, exhibiting a highly non-Langevin
behavior~\citep{Johnston2016,Herz2016}. This means that the ratio between the bimolecular
recombination rate $k_2\sim10^{-10}$ cm$^{3}$s$^{-1}$ and the charge carrier mobility
$\mu\sim10$~cm$^2$V$^{-1}$s$^{-1}$ is four orders of magnitude lower than the value
$e/(\epsilon_0\epsilon_r)$ predicted by the Langevin model~\citep{Herz2016,Wehrenfennig2014}.
This effect explains the long charge carrier diffusion lengths ($\sim1\mu$m) measured for halide
perovskites~\citep{Herz2016}.

The band gap of CH$_3$NH$_3$PbI$_3$ can be tuned via chemical substitution~\citep{Herz2016}. The
replacement of the CH$_3$NH$_3^+$ cation with CH(NH$_2$)$_2^+$ (formamidinium), Cs$^+$ or Rb$^+$
leads to a change in the optical band gap over a range of 0.3~eV~\citep{Eperon2014, Filip2014-1};
    the substitution of Pb$^{2+}$ by Sn$^{2+}$ reduces the band gap by approximately 0.4~eV~\citep{Hao2014}
and  the replacement of I$^-$ by the smaller anions Cl$^-$ and Br$^-$ blue-shifts the optical band
gap by up to 0.7~eV~\citep{Comin2015,Noh2013}. In practice, mixed cation, mixed metal and mixed
halide perovskites are currently explored in order to optimize light absorption over the entire
visible spectrum, but also to improve the stability of perovskite solar cells~\citep{McMeekin2016,
Eperon2016,Saliba2016}.

Theoretical investigations of the fundamental structural, electronic and vibrational properties of
CH$_3$NH$_3$PbI$_3$ have played an important role in the interpretation of the experimental data
summarized above. In the following we discuss some of the theoretical results obtained by us and our
coworkers in the study of CH$_3$NH$_3$PbI$_3$.

\begin{subsection}{Basic Electronic Structure of CH$_3$NH$_3$PbI$_3$}

\begin{figure*}[ht!]
\begin{center}
\includegraphics[width=0.7\textwidth]{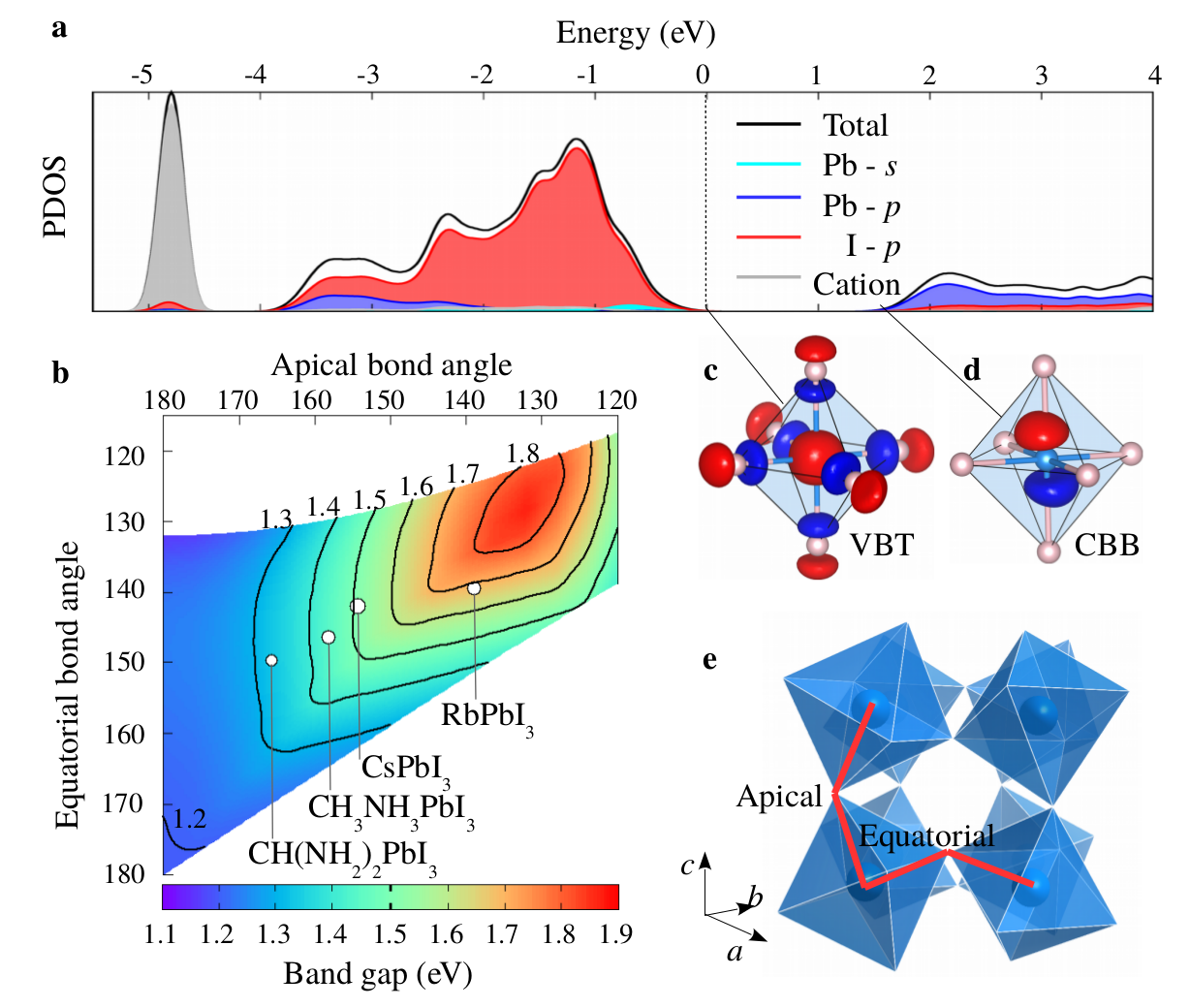}
\caption{{\bf Electronic structure and band gap tunability of CH$_3$NH$_3$PbI$_3$.}
{\bf a}. Projected density of states of CH$_3$NH$_3$PbI$_3$ calculated within the local density
approximation to density functional theory (DFT/LDA) without spin-orbit coupling for the
orthorhombic crystal structure. The bands are aligned to the valence band top (VBT) represented by
the dotted line. Adapted with permission from~\cite{Filip2014-1}, Copyright (2014) Nature Publishing 
Group.
{\bf b}. Two-dimensional map of the DFT band gap as a function of the apical and equatorial bond
angles. The band gaps for each point of the map were calculated for Platonic model structures, as
described by~\cite{Filip2014-1}, that do not contain the  A-site cation, as shown in {\bf e}. The
white disks represent the angular coordinates (apical and equatorial bond angles) of realistic
models of CH$_3$NH$_3$PbI$_3$, CH(NH$_2$)$_2$PbI$_3$, CsPbI$_3$ and RbPbI$_3$. 
Adapted with permission from~\cite{Filip2014-1}, Copyright (2014) Nature Publishing 
Group.
{\bf c - d}. Wave functions corresponding to the valence band top (c)  and the conduction band
bottom of a lead-iodide perovskite (d).
{\bf e}. Polyhedral representation of the Platonic model for the orthorhombic perovskite structure.
The apical and equatorial bond angles are represented by the red lines in the figure.\label{fig2}}
\end{center}
\end{figure*}

To construct the electronic band structure of CH$_3$NH$_3$PbI$_3$ it is necessary to know the
lattice parameters and  the atomic positions for the perovskite unit cell. For the low-temperature
orthorhombic phase, the unit cell was fully resolved for the first time by \cite{Baikie2013}.
The unit cell contains four PbI$_6$ octahedra and four CH$_3$NH$_3^+$ cations, 48 atoms in total,
arranged according to the $Pnma$ space group.

In the tetragonal and cubic phases, the unit cell contains four and one PbI$_6$ octahedra,
respectively, and the shape of the inorganic octahedral networks are well described
\citep{Stoumpos2013, Weller2015} by the space groups $I4/mcm$ and $Pm\bar{3}m$, respectively. At
variance with the inorganic scaffold, the organic cations have a disordered orientation inside the
perovskite cavity, in the tetragonal and cubic phases, and pose difficulties for any electronic
structure calculation that employs periodic boundary conditions. For example, structural relaxation
of the unit cell starting from arbitrary orientations of the CH$_3$NH$_3^+$ cations leads to
significant distortions of the unit cell and considerable changes to the electronic band structure
that are highly dependent on the choice of the orientation~\citep{Motta2015}. This effect is purely
an artefact of the chosen orientation of the cation, and bears no physical meaning.

To ensure that we can obtain meaningful results of the electronic properties of CH$_3$NH$_3$PbI$_3$
we have two options. One approach is to simulate the disordered orientation of the CH$_3$NH$_3^+$ by
averaging over its many possible orientations. This approach is best addressed using molecular
dynamics \citep{Carignano2015, Lahnsteiner2016} and requires a careful study of the size of the
supercell in order to correctly represent the disordered orientations of CH$_3$NH$_3^+$. The second
approach relies on the general property of ABX$_3$ perovskites by which the A-site cations do not
participate in bonds, and as a consequence they only contribute to electronic states far away from
the band edges \citep{Megaw}. In this latter approach, we can first calculate the electronic
properties of CH$_3$NH$_3$PbI$_3$ in the well known orthorhombic phase. Then we can investigate how
the change in the shape of the inorganic octahedral network affects these properties. In the
following, we describe the latter strategy.

In FIG.~\ref{fig2}a we show the projected density of states of CH$_3$NH$_3$PbI$_3$ in the low
temperature orthorhombic $Pnma$ case. The valence band top is occupied by I-$p$ and Pb-$s$
electrons, while the conduction band bottom is of I-$p$ and Pb-$p$ character. As expected, the
states localized on the CH$_3$NH$_3^+$ cation are located away from the band edges (approximately
5~eV). Based on FIG.~\ref{fig2}a we can construct a tight-binding argument to explain the
tunability of the band gap by  substitution of I anions or Pb cations. The replacement of I by Cl or
Br leads to an increase in the band gap, due to the lower-lying Br-$p$ and Cl-$p$ electronic states
that populate the valence band top; similarly, the substitution of Pb by Sn lowers the conduction
band bottom and reduces the band gap due to the lower-lying Sn-$p$ states. In addition, from FIG.
\ref{fig1}a, it is clear that the CH$_3$NH$_3^+$ cations do not contribute to the optical absorption,
and their role is to act as a structural filler and to ensure the charge neutrality of the unit cell.
In practice, the CH$_3$NH$_3^+$ cation can be replaced by a positive background charge, and the
electronic band structure would not change significantly~\citep{Filip2014-1}.

Taking advantage of this property, we can study how the shape of the cuboctahedral cavity influences
the electronic structure of the perovskite.~\cite{Filip2014-1} showed that the unit cell of the
perovskite lattice can be uniquely described by a `Platonic' model, whereby the PbI$_6$ octahedra
are ideal and identical throughout the lattice, and their orientation within the unit cell is
uniquely described by two angles: the equatorial and the apical bond angles (FIG.~\ref{fig2}e). FIG.
\ref{fig2}b depicts the variation of the band gap with the two bond angles of the `Platonic'
perovskite lattice. The gap increases as the bond angles decrease.

Two observations can be drawn from FIG.~\ref{fig2}b. Firstly, the size of the A-site cation in the
perovskite lattice influences the shape of the octahedral lattice, by changing the apical and
equatorial bond angles, which in turn enables the control of the band gap. This leads to the
following simple trend: the smaller the cation, the smaller the bond angles, the larger the band
gap. The band gap trend predicted in FIG.~\ref{fig2}b is in agreement with the trend observed
experimentally, and explains the band gap tunability with the substitution of the A-site cation
\citep{Eperon2014,Filip2014-1}. Secondly, based on FIG.~\ref{fig2}b we can explain the discontinuity
in the band gap dependence on temperature observed at the transition temperature of 160~K
\citep{Millot2015}. When the CH$_3$NH$_3$PbI$_3$ transitions from an orthorhombic to a tetragonal
phase, the apical bond angle increases from 160$^\circ$ to 180$^\circ$. According to the map in FIG.
\ref{fig2}b, the band gap should decrease by up to 0.2~eV, in good agreement with the sudden drop of
0.1~eV measured around 160~K.

The core  conclusion that can be drawn from FIG.~\ref{fig2}a is that the fundamental roles of the
organic cation CH$_3$NH$_3^+$ in the lead-iodide perovskite lattice are to balance the charge, and to
stabilize the shape of the perovskite lattice. Therefore, in order to ensure that we are studying a
physically meaningful crystal structure, in the following discussions we focus on the
low-temperature orthorhombic $Pnma$ phase, as refined by~\cite{Baikie2013}.

\end{subsection}

\begin{subsection}{The Quasiparticle Band Structure of CH$_3$NH$_3$PbI$_3$}

\begin{figure*}[ht!]
\begin{center}
\includegraphics[width=0.8\textwidth]{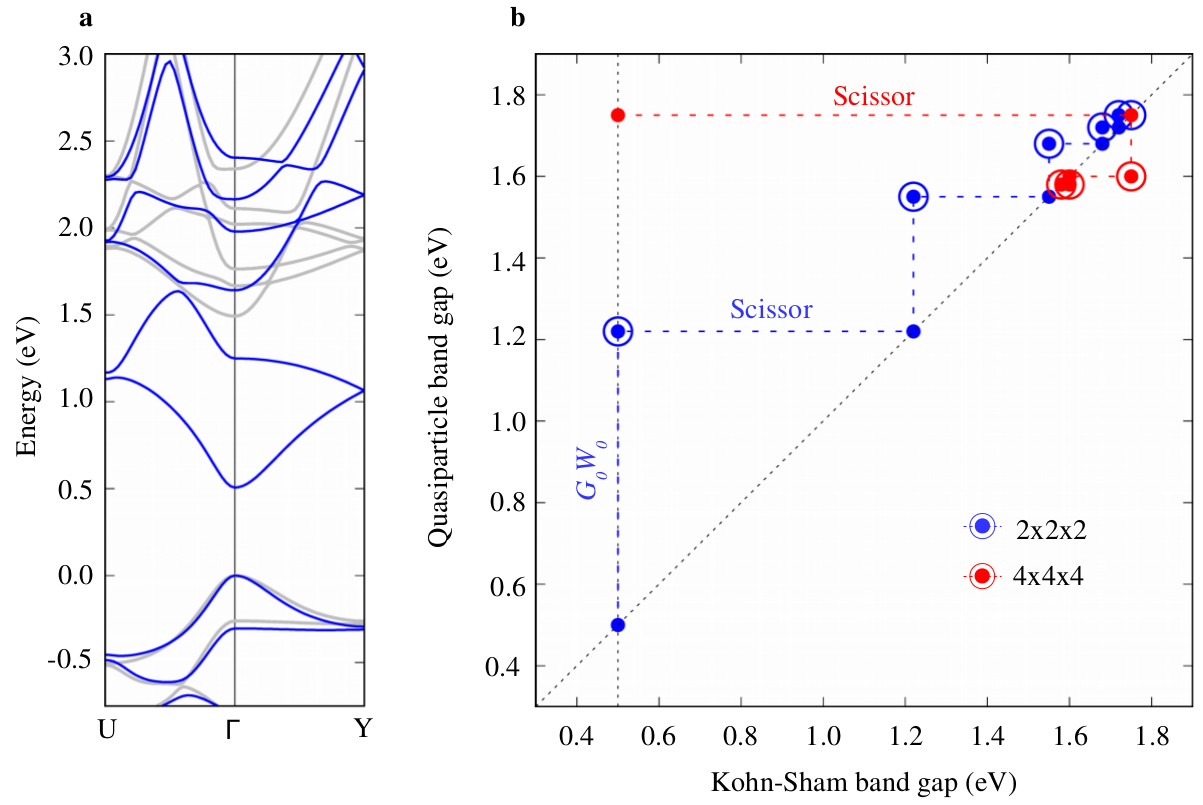}
\caption{{\bf Band structure of CH$_3$NH$_3$PbI$_3$.}
{\bf a}. DFT/LDA band structure of CH$_3$NH$_3$PbI$_3$ calculated with (blue line) and without (grey line)
including spin-orbit coupling effects. The calculations were performed for the orthorhombic $Pnma$
phase using norm-conserving Troullier Martins pseudopotentials as described by~\cite{Filip2014-2}.
{\bf b}. Convergence of the self-consistent scissor $GW$ scheme for the quasiparticle band gap of
CH$_3$NH$_3$PbI$_3$. The self-consistent scissor $GW$ scheme is described in detail by
\cite{Filip2014-2}. For all  quasiparticle calculations we have used 1000 bands, and plane wave
cutoffs of 35~Ry and 6 Ry for the exchange and polarizability respectively. The diagram shows the
self-consistent scissor scheme calculated for a $\Gamma$-centered $2\times2\times2$ {\bf k}-point
grid (blue) and a $4\times4\times4$ {\bf k}-point grid (red). All quasiparticle calculations include
semicore $d$ electrons for both Pb and I, as discussed by~\cite{Filip2014-2}.\label{fig3}}
\end{center}
\end{figure*}

The calculations shown so far in FIG.~\ref{fig2} were obtained within DFT in the scalar-relativistic
approximation. However, the presence of Pb in CH$_3$NH$_3$PbI$_3$ requires a relativistic treatment.
In FIG.~\ref{fig3}a we show that the spin-orbit coupling effect is crucial for the correct
calculation of the band edge topology. Not only relativistic effects lower the conduction band
bottom of CH$_3$NH$_3$PbI$_3$ by approximately 1~eV~\citep{Even2013}, but also they change the shape
of the band edges. In particular, the effective masses become fully isotropic upon inclusion of
spin-orbit coupling, by contrast with those calculated in the scalar-relativistic approach
\citep{Menendez2014}. As shown in Table~\ref{tb1}, fully-relativistic DFT calculations underestimate
the band gap by $\sim$1~eV and the reduced effective mass by a factor of two.

The underestimation of the band gap and effective masses of DFT can be corrected by including
quasiparticle effects. The quasiparticle band gap and band structure can be calculated within the
$GW$ approximation~\citep{Hedin}. The quasiparticle eigenvalues $E_{n\mathbf{k}}$ are calculated
starting from the single-particle DFT/LDA eigenvalues $\epsilon_{n\mathbf{k}}$ as~\citep{Hedin,
Hybertsen}:

\begin{equation}
E_{n\mathbf{k}} = \epsilon_{n\mathbf{k}} + Z(\epsilon_{n\mathbf{k}}) \langle n\mathbf{k} |
\hat{\Sigma}(\epsilon_{n\mathbf{k}}) - V_{xc} | n\mathbf{k} \rangle,
\label{eq1}
\end{equation}

where $|n\mathbf{k}\rangle$ denotes a single-particle eigenstate with band index $n$ and crystal
momentum $\textbf{k}$, $Z(\omega) = (1-\partial {\rm Re}\hat{\Sigma}/\partial \omega)^{-1}$ is the
quasiparticle renormalization, $V_{xc}$ is the exchange-correlation potential, and $\hat{\Sigma}
= iG_0W_0$ is the quasiparticle self-energy~\citep{Hybertsen}. In the expression of $\hat{\Sigma}$,
$G_0$ is the non-interacting Green's function calculated using single-particle eigenvalues and
eigenstates as obtained from DFT. The screened Coulomb interaction has the expression $W_0 =
\epsilon^{-1}v$, where $v$ is the bare Coulomb interaction and $\epsilon^{-1}$ is the inverse
dielectric matrix.

In the case of CH$_3$NH$_3$PbI$_3$ it has been shown that the one-shot $G_0W_0$ quasiparticle band
gap is underestimated with respect to experiment by up to 0.4~eV~\citep{Brivio2014,Filip2014-2}, as
show in Table~\ref{tb1}. This is primarily due to the underestimated single particle band gap
calculated at the DFT level, which leads to an overestimation of the dielectric screening in
$W_0$~\citep{Brivio2014,Filip2014-2}. The quasiparticle band gap underestimation at the $G_0W_0$
level appears irrespective of the exchange-correlation functional employed for the DFT starting
point, and it tends to worsen when more core electrons are included in the valence configuration of
I~\citep{Scherpelz2016}. However, this sensitivity to the DFT starting point can be mitigated by
calculating the electron self-energy self-consistently~\citep{Schilfe2006}.

\begin{table*}[t!]
\begin{center}
\begin{tabular}{  l c c c c c c c l }
\hline
\hline
 & & DFT & & $G_0W_0$ & & $GW$ & & Exp.\\
\hline
Band gap (eV)                  && 0.50$^{a,b,c}$ &&  1.20$^{a,b,c}$ && 1.57$^{c}$ && 1.65$^{c,d}$\\
\hline
    $\epsilon_{\infty}$            && \multicolumn{5}{c}{5.8$^{b,e}$}    && 6.5$^f$  \\
    $\epsilon_0$                   && \multicolumn{5}{c}{25.3$^e$}   && 30.5$^g$\\
\hline
Hole effective mass ($m_{\rm e}$)     && 0.14$^{b,c}$ &&  0.18$^{b,c}$ && 0.23$^{b,c}$ && N/A \\
Electron effective mass ($m_{\rm e}$) && 0.12$^{b,c}$ &&  0.16$^{b,c}$ && 0.22$^{b,c}$ && N/A \\
Reduced effective mass ($m_{\rm e}$)  && 0.06$^{b,c}$ &&  0.09$^{b,c}$ && 0.11$^{b,c}$ && 0.104$
										\pm$0.003$^h$\\
\hline
Exciton binding energy (meV)    && \multicolumn{5}{c}{44} && 16$^h$  \\
\hline
\hline
\end{tabular}
\caption{Calculated values of the band gap, high and low-frequency dielectric constants, electron,
hole and reduced  effective masses within DFT/LDA and within the $GW$ approximation including the
self-consistent scissor correction of ~\cite{Filip2014-2}. The band gaps are expressed in eV and the
effective masses are expressed in the unit of electron masses ($m_e$). The reduced effective mass is
calculated using the expression: ${\rm\mu}^* = \frac{m_{\rm h}^*m_{\rm e}^*}{m_{\rm h}^*+
m_{\rm e}^*}$, where $m_{\rm e}^*$,  $m_{\rm h}^*$ and $\rm{\mu}^*$ are the electron, hole and
reduced effective masses respectively. The exciton binding energy is calculated using  the
expression $E_{\rm b} = \rm{\mu}^*/\epsilon_{\infty}^2 \times E_{\rm Ry} $~eV, with $E_{\rm Ry} =
13.60565$~eV, the reduced effective mass $\mu^*=0.11$~$m_{\rm e}$ and the high-frequency dielectric
constant $\epsilon_{\infty}  = 5.8$.\\~\label{tb1}
$^a$~\cite{Filip2014-2}\\ $^b$~\cite{Filip2015}\\ $^c$~\cite{Davies2017}\\ $^d$~\cite{Millot2015}\\
    $^e$~\cite{Perez2015}\\ $^f$~\cite{Hirasawa1994}\\ $^g$~\cite{Poglitsch1987}\\ $^h$~\cite{Myiata2015}}
\end{center}
\end{table*}

FIG.~\ref{fig3}b shows an approximate self-consistency scheme introduced by \cite{Kioupakis2008}
and applied to the case of CH$_3$NH$_3$PbI$_3$ by \cite{Filip2014-2}, the self-consistent scissor
$GW$. In this approach the $G_0W_0$ band gap correction $\Delta$ is first calculated, and then a
scissor correction of magnitude $\Delta$ is applied to all eigenstates in the conduction band. These
two steps are repeated until the band gap is converged, as shown in FIG.~\ref{fig3}b
\citep{Kioupakis2008, Filip2014-2}. The final converged band gap is of 1.57~eV, within 80~meV of the
measured optical band gap of 1.65~eV~\citep{Millot2015, Davies2017}. This small difference is
accounted for by the absence of electron-hole interaction effects, and electron-phonon effects.

FIG.~\ref{fig4}a shows the comparison between the $GW$ quasiparticle band structure and the DFT/LDA
band structure obtained from Wannier interpolation~\citep{Marzari1997, Souza2001}. Qualitatively the
band dispersions calculated within DFT/LDA and $GW$ are similar, with the quasiparticle valence band
exhibiting a noticeably more pronounced curvature. A more quantitative comparison between the
DFT/LDA and $GW$ band dispersions is given by the isotropic effective masses, shown in Table
\ref{tb1}. The effective mass tensor is calculated as the inverse of the second derivatives of the
energy with respect to the crystal momentum,
$m_{ij}^*=\hbar^2(\partial^2\epsilon/\partial k_i \partial k_j)^{-1}$, and the values reported in
Table~\ref{tb1} are the averages of the eigenvalues of the effective mass tensors. From Table
\ref{tb1} we can see that the DFT/LDA and $GW$ effective masses differ by almost a factor of two.
In addition, the $GW$ reduced effective mass of 0.11~$m_e$ is in excellent agreement with the
experimental value of 0.10~$m_e$ reported by~\cite{Myiata2015}, thus establishing the accuracy of
the quasiparticle band structure calculations.

An important consequence of the quasiparticle effects in the calculation of the band structure of
CH$_3$NH$_3$PbI$_3$ is the band edge parabolicity. While both DFT/LDA and $GW$ yield isotropic band
edges, the DFT/LDA bands are parabolic only in a narrow energy range near the band edges. On the
other hand, the quasiparticle valence and conduction band edges exhibit a parabolic profile over a
range of 300~meV (FIG.~\ref{fig4}b). This observation is most clearly visualized in the joint
density of states (JDOS), shown in FIG.\ref{fig4}b, where the parabolic shape of the bands extends
over approximately 0.3~eV for the quasiparticle JDOS, while the JDOS calculated from DFT/LDA departs
from the parabolic shape within 50~meV from the onset.

\begin{figure*}[t!]
\begin{center}
\includegraphics[width=0.9\textwidth]{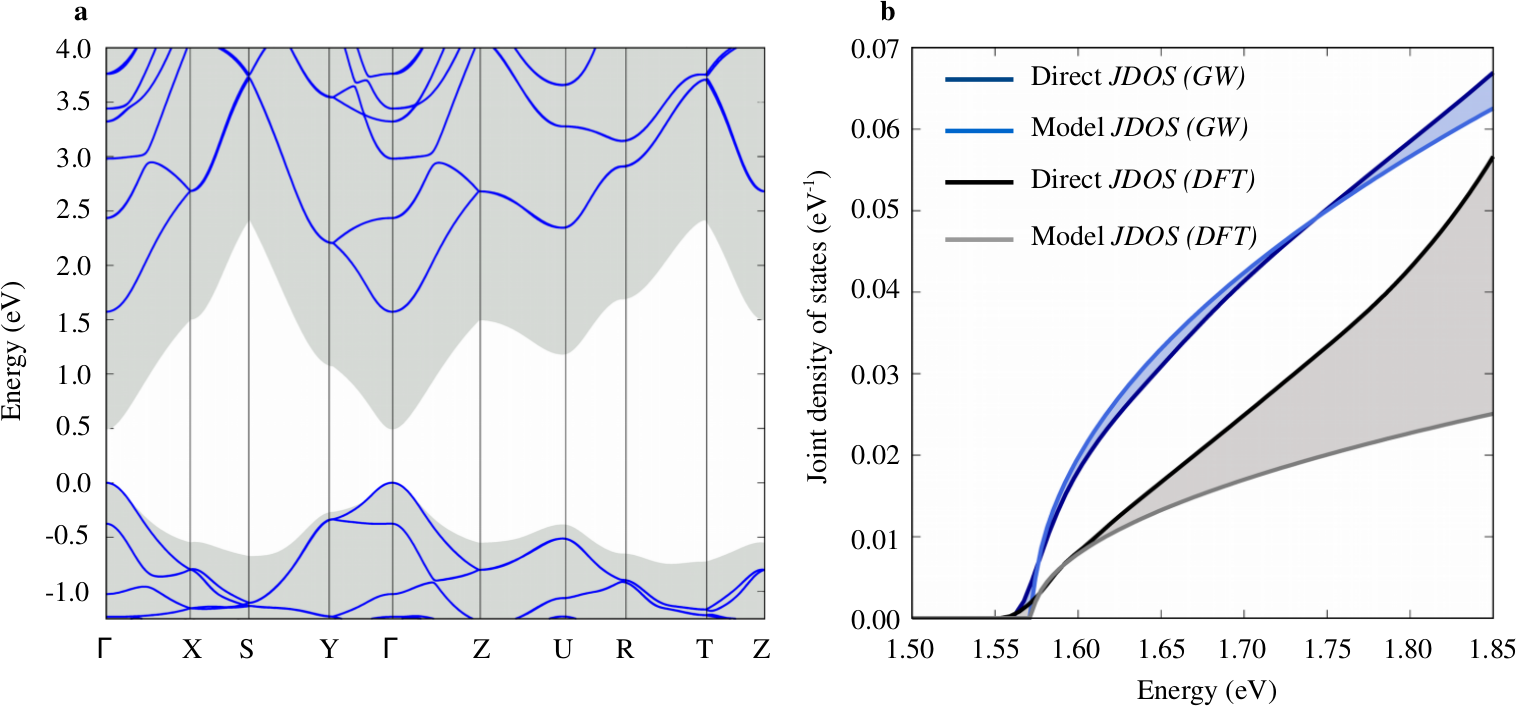}
\caption{{\bf Band parabolicity.}
{\bf a}. Quasiparticle band structure of CH$_3$NH$_3$PbI$_3$. The quasiparticle band structure is
represented by the blue lines, the grey background is the DFT/LDA band structure. The quasiparticle
band structure was obtained by performing a Wannier interpolation of the quasiparticle eigenvalues
calculated for a $4\times4\times4$ $\Gamma$-centered grid, using the $GW$ and the self-consistent
scissor scheme~\citep{Filip2015, Davies2017}.
{\bf b}. Joint density of states of CH$_3$NH$_3$PbI$_3$ calculated within DFT (black line) and $GW$
(dark blue line). The direct joint densities of states were calculated using the definition
$J(\omega) = \sum_{cv\textbf{k}} \delta(\epsilon_{c\textbf{k}}-\epsilon_{v\textbf{k}}-\omega)$,
where $c$ and $v$ are the band indices for the conduction and valence band states respectively, and
$\textbf{k}$ is the crystal momentum. The summation over the crystal momenta is performed by
discretizing the Brillouin zone on  a $\Gamma$-centered grid of $40\times40\times40$ points. The DFT
and $GW$ eigenvalues are obtained on this grid from Wannier interpolation~\citep{Filip2015,
Davies2017}. The model joint densities of states (light blue line for $GW$ and grey for DFT) are
calculated using the assumption that the bands are parabolic using,
$J^{{\rm DFT}/GW}(\omega) = 1/4\pi^2(2\mu^{{\rm DFT}/GW}/\hbar^2)^{3/2}(\omega-E_g)^{1/2}$, where
$\mu$ is the reduced effective mass calculated from DFT (0.06 $m_{\rm e}$) or $GW$ (0.11
$m_{\rm e}$). Both the model and the direct JDOS calculated from DFT were rigidly shifted in order
to match the $GW$ onset, for clarity. ~\label{fig4}}
\end{center}
\end{figure*}
The parabolicity of the conduction and valence band edges has important implications in the
understanding of both absorption and recombination mechanisms in CH$_3$NH$_3$PbI$_3$. Using a
combined experimental and theoretical analysis of the optical absorption lineshape,
\cite{Davies2017} showed that the optical absorption spectrum of CH$_3$NH$_3$PbI$_3$ can be modelled
using the Elliot's theory~\citep{Elliot} up to 100~meV above the absorption onset. Within this
premise,~\cite{Davies2017} decoupled the excitonic and continuum part of the absorption spectrum,
and calculated the bimolecular recombination rate using the van Roosbroek-Shockley formalism
\citep{Roosbroek}. By comparing this model to direct transient spectroscopic measurements,
\cite{Davies2017} demonstrated that bimolecular recombination in CH$_3$NH$_3$PbI$_3$, is an inverse
absorption process. This result reinforces the conclusion that CH$_3$NH$_3$PbI$_3$ is in many
respects very  similar to conventional semiconductors, such as GaAs~\citep{Herz2016}.

\end{subsection}

\begin{subsection}{Vibrational Properties of CH$_3$NH$_3$PbI$_3$}

The vibrational spectrum is a key piece of information in the study of the optoelectronic properties
of CH$_3$NH$_3$PbI$_3$. Lattice vibrations can be analyzed experimentally from IR~\citep{Perez2015}
and Raman~\citep{Ledinsky2015} spectroscopy, as well as inelastic neutron scattering
\citep{Druzbicki2016}, and can be directly compared with {\it ab initio} calculations. Compared to
the latter two techniques, IR spectroscopy is most accessible, as high-resolution measurements can be
achieved in a short time frame over a wide range of frequencies
\citep{Perez2015}.

\begin{figure*}[t!]
\begin{center}
\includegraphics[width=0.9\textwidth]{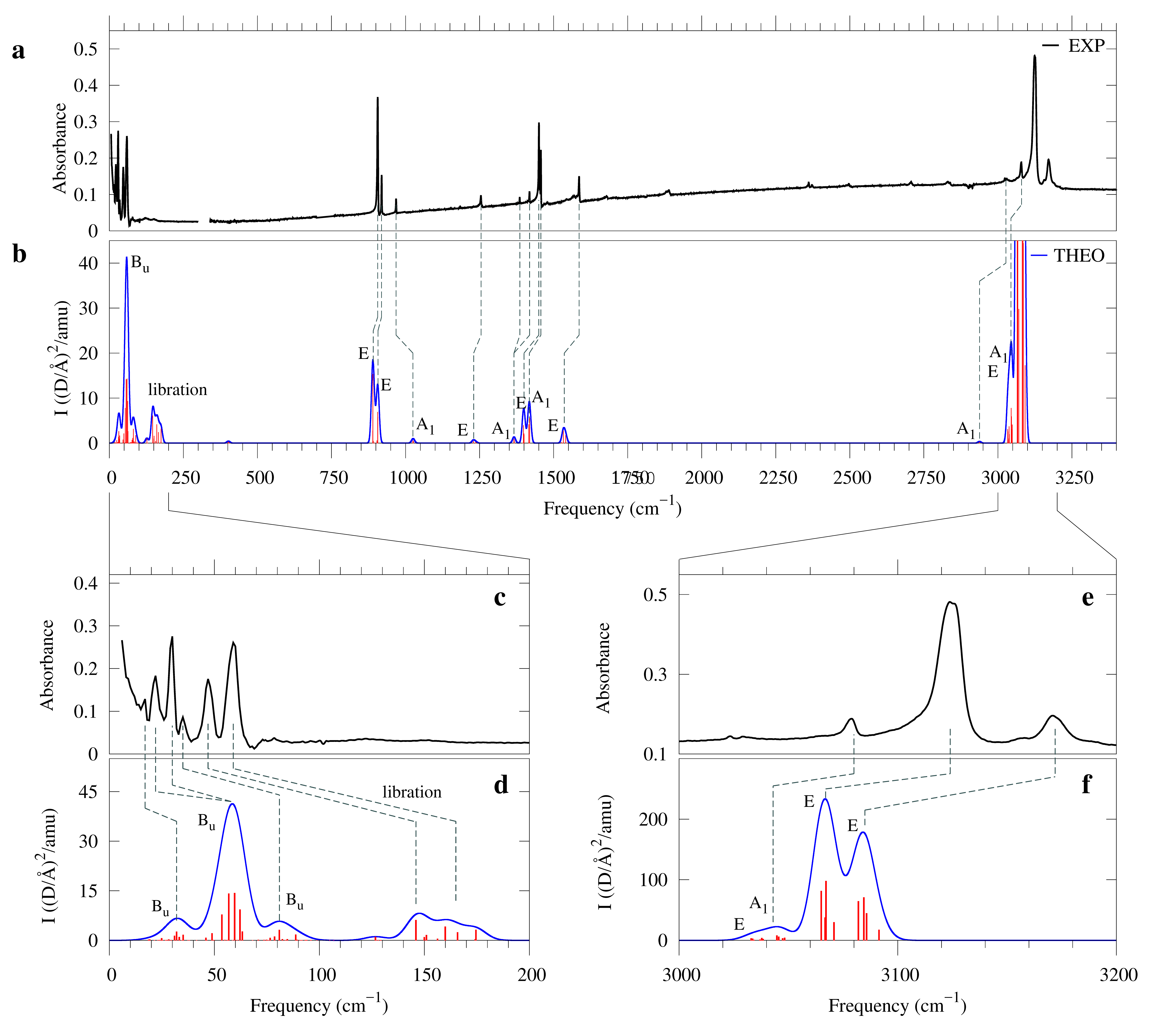}
\caption{{\bf Experimental and theoretical IR spectra of CH$_3$NH$_3$PbI$_3$ at low temperature.}
{\bf a}. IR spectrum of CH$_3$NH$_3$ measured on thin film at 10~K. The dashed lines indicate the
assignment of the most prominent peaks of the spectrum to features in the calculated IR spectrum in
{\bf b}. The IR spectrum is measured at normal incidence, and only TO normal modes can be excited in
this geometry.
{\bf b}. Calculated IR spectrum of CH$_3$NH$_3$PbI$_3$ in the low temperature orthorhombic
structure. The red sticks are the calculated infra-red intensities and the blue line corresponds
to the spectrum calculated by representing each absorption peak with a Gaussian broadening. To be
able to compare with the experimental data obtained within the geometry described in {\bf a}, only
TO modes are included in the calculated IR spectrum, as discussed by~\cite{Perez2015}.
{\bf c, d}. Close-up of the comparison between the experimental ({\bf c}) and theoretical ({\bf d})
IR spectra in the low frequency region. {\bf e, f}. Close-up of the comparison between experimental
({\bf e}) and theoretical ({\bf f}) IR spectra in the high-frequency region. Reprinted
with permission from~\cite{Perez2015} \label{fig5}, Copyright (2015) American Chemical Society.
}
\end{center}
\end{figure*}

The vibrational properties of CH$_3$NH$_3$PbI$_3$ can be calculated within density functional
perturbation theory (DFPT). The vibrational spectrum of CH$_3$NH$_3$PbI$_3$ in the orthorhombic
$Pnma$ phase consists of 144 normal modes at the $\Gamma$ point, with frequencies extending up to
3100 cm$^{-1}$~\citep{Perez2015}. The vibrational modes of CH$_3$NH$_3$PbI$_3$ can be classified
into three main categories: vibrations of the CH$_3$NH$_3^+$ cations, vibrations of the inorganic
PbI$_3$ network, and mixed modes. In addition, the vibrations of the CH$_3$NH$_3^+$ cations can be
decomposed into rigid translations, spinning vibrations around the C-N axis, librations around the
center of mass of the cation, and internal vibrations. Up to 60~cm$^{-1}$, the main contribution to
the vibrational phonon modes comes from the internal  vibrations of the inorganic network, while
above 60~cm$^{-1}$ the vibrational phonon modes are almost exclusively due to the vibrations of the
organic cations~\citep{Perez2015}. \cite{Perez2015} describes the precise contribution of these
three categories of vibrations to each of the vibrational modes. These contributions were calculated
by a systematic decomposition of each mode followed by a factor-group analysis of their symmetries.

Using the calculated vibrational frequencies of CH$_3$NH$_3$PbI$_3$, the IR spectrum is obtained by
calculating the absorption spectrum in the IR frequency range as~\citep{Perez2015}:

\begin{equation}
    I(\omega) = \frac{e^2}{M_0}\sum_{\alpha \nu} |Z_{\alpha \nu}^*|^2 \delta (\omega-\omega_\nu),
\end{equation}

where $M_0$ is the average mass over  the unit cell, $e$ is the electronic charge, $\omega_\nu$ are
the vibrational eigenfrequencies at the zone center and
$\displaystyle{Z_{\alpha \nu}^* = \sum_{\kappa \beta} \sqrt{\frac{M_0} {M_\kappa}}
e_{\kappa \beta,\nu}Z_{\kappa,\alpha\beta}^*}$, with $M_\kappa$ being the nuclear mass of atom
$\kappa$ in the unit cell, $e_{\kappa \beta,\nu}$ the vibrational eigenmode of atom $\kappa$ in the
Cartesian direction $\beta$, with vibrational frequency $\omega_\nu$, and $Z_{\kappa,\alpha\beta}^*$
is the Born effective charge~\citep{Perez2015}.

In FIG.~\ref{fig5} we show a comparison between the experimental and theoretical IR spectra of
CH$_3$NH$_3$PbI$_3$, as measured and calculated by~\cite{Perez2015}. Each of the features in the
experimental IR spectrum is assigned to a peak and a vibrational symmetry in the calculated spectrum.
The theoretical and experimental spectra agree well for almost the entire frequency range. The two
main discrepancies are shown in the FIG..~\ref{fig5}c-f. At high frequency, the three peaks are
redshifted by approximately 70~cm$^{-1}$. This discrepancy is shown by~\cite{Perez2015} to be
corrected by the use of the generalized gradient approximation (GGA) instead of LDA for the
calculation of the exchange-correlation potential in DFPT. In the low frequency (FIG.~\ref{fig5}c-d)
region of the spectrum, DFPT calculations appear to overestimate the vibrational frequencies of the
librational modes of CH$_3$NH$_3^+$ around 150~cm$^{-1}$. This discrepancy (FIG.~\ref{fig5}e-f) is
consistent with other studies in literature~\citep{Mosconi2014, Druzbicki2016}, and it appears
regardless of the choice of exchange-correlation functional (LDA vs. GGA), inclusion of relativistic
spin-orbit coupling effects, van der Waals interactions or anharmonic effects~\citep{Perez2017}.
Moreover, \cite{Perez2017} show that the calculated vibrational density of states is in  good
agreement with inelastic neutron scattering spectra~\citep{Druzbicki2016}, suggesting that the
discrepancy between the experimental and calculated IR spectra may be related to a mismatch between
the experimental and theoretical IR intensities.

Having established both the electronic and vibrational properties of CH$_3$NH$_3$PbI$_3$, we have
access to the high  and low frequency dielectric permitivities, $\epsilon_\infty$ and $\epsilon_0$,
respectively. The high-frequency permitivity does not depend on lattice vibrations, but its value
does depend on the calculated band gap. For this reason, the closest agreement with the experimental
value $\varepsilon_\infty = 6.5$~\citep{Hirasawa1994} is obtained from finite-electric field
calculations within scalar-relativistic DFT~\citep{Perez2015, Filip2015} (5.8, see Table~\ref{tb1}).
The low-frequency permitivity depends on the vibrational frequencies of CH$_3$NH$_3$PbI$_3$, with
the vibrations of the inorganic PbI$_6$ network carrying the largest contribution~\citep{Perez2015}.
\cite{Perez2015} reports a value of $\varepsilon_0=25.3$, as calculated from DFPT, in very good
agreement with the experimental low-frequency dielectric constant $\varepsilon_0=30.5$
\citep{Poglitsch1987}. As shown by~\cite{Perez2015}, the large difference between the high- and
low-frequency dielectric permitivities is due almost entirely to the vibrations of the inorganic
PbI$_6$ network.

\end{subsection}

\begin{subsection}{Electron-Phonon Coupling in CH$_3$NH$_3$PbI$_3$}

\begin{figure*}[t!]
\begin{center}
\includegraphics[width=0.7\textwidth]{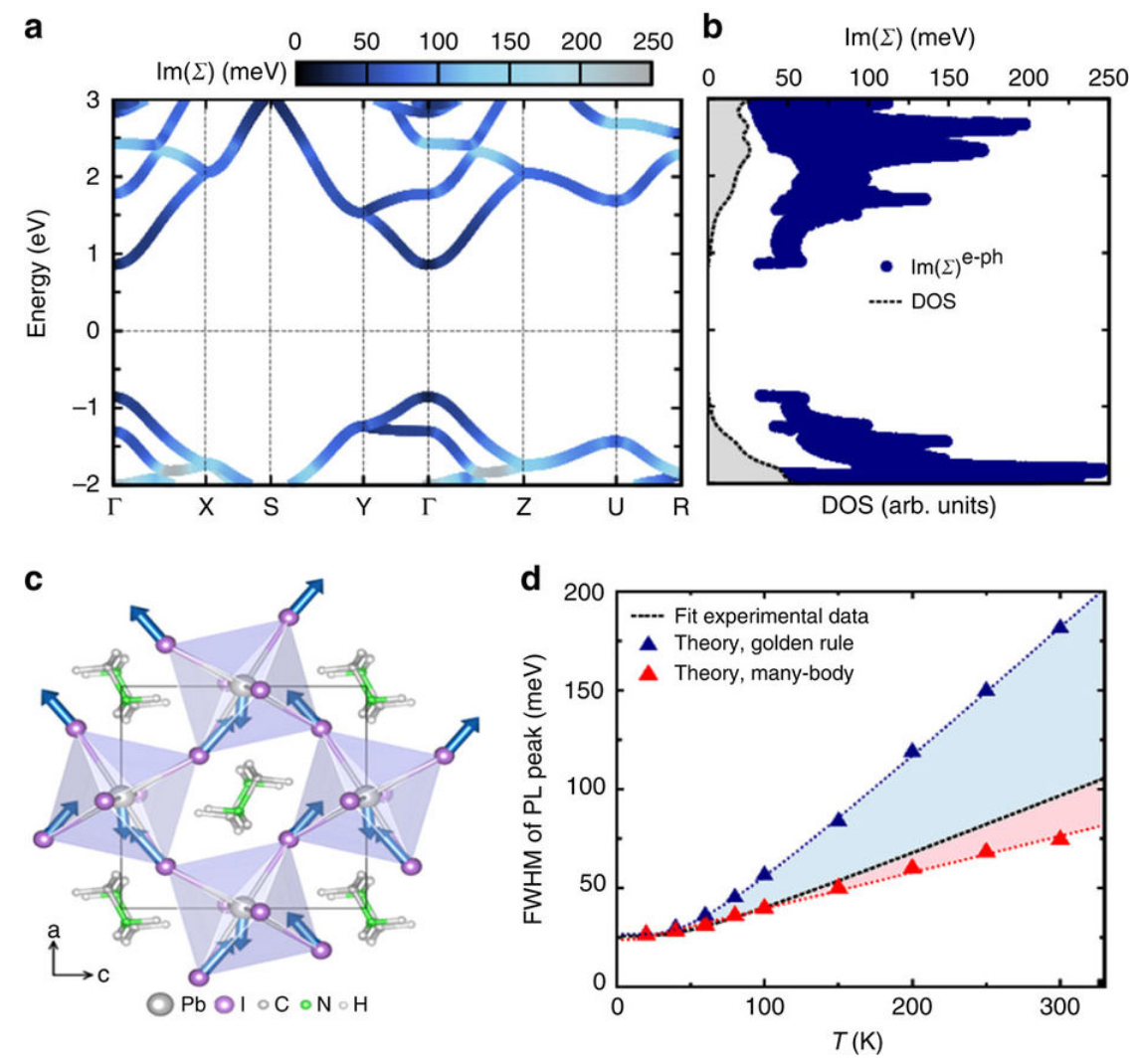}
\caption{{\bf Electron-phonon coupling and PL broadening of CH$_3$NH$_3$PbI$_3$.}
{\bf a}. $GW$ quasiparticle band structure of the orthorhombic CH$_3$NH$_3$PbI$_3$, combined with
the heat map of the imaginary part of the electron-phonon self-energy, ${\rm Im}(\Sigma)$ at T =
200~K. The zero of the energy is placed in the middle of the band gap.
{\bf b}. Imaginary part of the electron-phonon self-energy (dark blue line) and the electronic
density of states (dotted black line). $2{\rm Im}(\Sigma)$ corresponds to the electron-hole
linewidth arising from electron-phonon coupling (without the quasiparticle renormalization factor
$Z$).
{\bf c}. Ball-and-stick representation of the LO vibration responsible for the broadening of the PL
peaks.
{\bf d}. Temperature dependence of the full width at half maximum (FWHM) of the PL peak in
CH$_3$NH$_3$PbI$_3$. The dashed black line corresponds to the fit to experimental data and the blue
and red triangles correspond to theoretical calculations using Fermi's golden rule and the Brillouin
Wigner perturbation theory (red triangles), respectively. The theoretical broadening is obtained as
the sum of  $2{\rm Im}(\Sigma)$ at the valence and conduction band edges, rigidly shifted by a
constant inhomogeneous  broadening of 26~meV. Reprinted with permission from~\cite{Wright2016}, 
Copyright (2016) Nature Publishing Group.
 \label{fig6}
}.
\end{center}
\end{figure*}

Understanding the interaction between electrons and lattice vibrations is essential in the
investigation of the transport properties and charge carrier recombinations in CH$_3$NH$_3$PbI$_3$.
In particular, electron-phonon coupling manifests itself prominently in the temperature dependent
charge carrier mobility and in the broadening of the photoluminescence spectra~\citep{Johnston2016}.
The charge carrier mobility decreases with increasing temperature as $T^{m}$, where $m$ takes values
between -1.4 and -1.6~\citep{Brenner2015, Wright2016}. Such behaviour is typical of non-polar
semiconductors such as silicon or germanium, where electron-phonon interactions at room temperature
are almost exclusively governed by the interaction with acoustic phonons. It is, however, unexpected
in the case of CH$_3$NH$_3$PbI$_3$, given the polar nature of the this compound.

As shown by \cite{Wright2016}, the nature of electron-phonon interactions in CH$_3$NH$_3$PbI$_3$ can
be investigated by analyzing the photoluminescence linewidth as a function of temperature. In their
study,~\cite{Wright2016} find that the contribution of acoustic phonons to the broadening of the
photoluminescence linewidth is negligible compared to that of the longitudinal-optical (LO) modes.
This argument is further corroborated by first principles calculations of the electron-phonon
coupling, shown in FIG.~\ref{fig6}~\citep{Wright2016}.

The photoluminescence linewidth can be directly associated with the imaginary part of the
electron-phonon self energy, calculated using the following expression~\citep{Wright2016}:

    \begin{widetext}
\begin{equation}
\Sigma_{n\mathbf{k}} = \sum_{m\nu\mathbf{q}}|g_{mn}^\nu(\mathbf{k},\mathbf{q})|^2\Big[
\frac{n_{\mathbf{q},\nu}+f_{m\mathbf{k}+\mathbf{q}}}{\varepsilon_{n\mathbf{k}}-
\varepsilon_{m\mathbf{k}+\mathbf{q}}-\omega_{\mathbf{q}\nu}-i\eta}
+\frac{n_{\mathbf{q},\nu}+1-f_{m\mathbf{k}+\mathbf{q}}}{\varepsilon_{n\mathbf{k}}-
    \varepsilon_{m\mathbf{k}+\mathbf{q}}-\omega_{\mathbf{q}\nu}-i\eta}\Big],
\end{equation}
    \end{widetext}

where $f_{m\mathbf{k}+\mathbf{q}}$ and $n_{\mathbf{q},\nu}$ are the Fermi-Dirac and Bose-Einstein
occupations, $\epsilon_{n\mathbf{k}}$ and $\hbar\omega_{\mathbf{q}\nu}$ are the electron and phonon
energies respectively, $\eta$ is a small broadening parameter, and
$g_{mn}^\nu(\mathbf{k},\mathbf{q})$ is the electron-phonon matrix element calculated by taking into
account only the interaction with the LO phonons, as discussed by~\cite{Verdi2015}.

FIG.~\ref{fig6}a shows how the imaginary part of the electron and hole linewidth are distributed
across the electronic states in the quasiparticle band structure~\citep{Wright2016}. The linewidth
increases as the density of states increases (FIG.~\ref{fig6}b), because more states become
available for electronic transitions. The predominant contribution to the electron-phonon
self-energy comes from the LO phonon mode with an energy of 13~meV, which corresponds to the
vibrations shown in FIG.~\ref{fig6}c. This value is in excellent agreement with the dominant phonon
energy of 11~meV determined from the experiments~\citep{Wright2016}. FIG.~\ref{fig6}d shows that the
measured and the calculated trend for the PL broadening as a function of temperature are in very
good agreement, thereby confirming that the predominant electron-phonon interaction is that between
charge carriers and longitudinal optical phonons. In addition, both theoretical and experimental
results reported by~\cite{Wright2016} show that CH$_3$NH$_3$PbBr$_3$ exhibits an electron-phonon
coupling which is 40\% stronger than for the iodide, pointing to potential directions for
controlling such effects through chemical substitution.

\end{subsection}

\begin{subsection}{Desirable electronic structure properties and the basis for materials design}

Experimental and theoretical studies of the electronic and optical properties of CH$_3$NH$_3$PbI$_3$ 
have so far elucidated many of its fundamental properties. Perhaps the most striking finding is that 
CH$_3$NH$_3$PbI$_3$ exhibits optoelectronic properties which are remarkably similar to the best 
inorganic semiconductors such as GaAs. Despite much progress on the understanding of the fundamental 
properties of CH$_3$NH$_3$PbI$_3$ and related compounds, several important questions remain to be 
addressed. For example it would be important to investigate charge carrier transport and 
recombination processes~\citep{Herz2016}, degradation mechanisms~\citep{Leijtens2017}, and defect 
physics~\citep{Ball2017}. All these topics are active areas of current research in this field.

Much of the current success of halide perovskites owes to the continued in-depth 
investigation of their fundamental optoelectronic properties, but also to the development of 
practical strategies to control these properties through chemical substitution and to design 
new materials with targeted functionalities. For example, chemical substitution has been 
systematically explored as a means to tune the optical absorption properties~\citep{Filip2014-1, 
Eperon2016}, improve the stability~\citep{McMeekin2016} and reduce the toxicity
\citep{Giustino2016} of lead-halide perovskites. 

In particular, there is currently an ongoing search for alternative lead-free halide perovskites 
which retain the optoelectronic properties of CH$_3$NH$_3$PbI$_3$, and remove any potential 
environmental concerns due to the presence of lead. The nearly ideal optoelectronic properties of 
lead-halide perovskites set a very high bar for potential lead-free candidates: new compounds must 
exhibit a band gap in the visible range, long carrier lifetimes, good charge carrier mobilities, low 
exciton binding energies, and shallow or electrically-inactive defects. In the next section we 
review our efforts in the search for novel halide perovskites that might exhibit this unique 
combination of optoelectronic properties.

\end{subsection}
\end{section}

\begin{section}{Design of lead-free perovskites}

The presence of lead in lead-halide perovskites has raised questions on the environmental impact of 
perovskite solar cells. While the content of lead by weight in these devices is well below the 
limits set by governmental agencies, the prospect of developing lead-free perovskites is very 
attractive. Until now a lead-free material that can compete with CH$_3$NH$_3$PbI$_3$ and related 
compounds has not been found, but substantial progress has been made in the development of new 
lead-free perovskites.

Among the different approaches to develop new lead-free perovskites, first-principles computational 
design has emerged as a powerful tool for screening new materials and identifying promising 
candidates. In fact, over the last two years several new lead-free halide double perovskites have 
been first proposed by theoretical studies, and subsequently synthesized in the lab. In the next 
sections representative examples in this area are discussed.

 \begin{subsection}{Homovalent Pb replacement}
In order to replace lead in the APbX$_3$ structure, the simplest starting point is to search for 
alternative cations in the +2 oxidation state. Hypothetical new perovskites should be stable towards 
air, moisture, illumination, and heat, and should exhibit band gap and carrier effective masses 
similar to those of CH$_3$NH$_3$PbI$_3$. The most obvious choice would be to replace Pb$^{2+}$ with 
another atom of Group~IV, such as Sn$^{2+}$. Indeed Sn-based perovskites were shown to exhibit ideal 
band gaps for photovoltaics applications, as well as good charge carrier mobilities
\citep{Stoumpos2013}. However, perovskites containing Sn$^{2+}$ tend do degrade rapidly due to the 
oxidation of the cation to Sn$^{4+}$. This limitation was recently overcome by developing mixed 
Sn/Pb halide perovskites, and the first applications to tandem device architectures are very 
promising \citep{McMeekin2016,Eperon2016}.

\begin{figure*}[ht!]
\begin{center}
\includegraphics[width=0.9\textwidth]{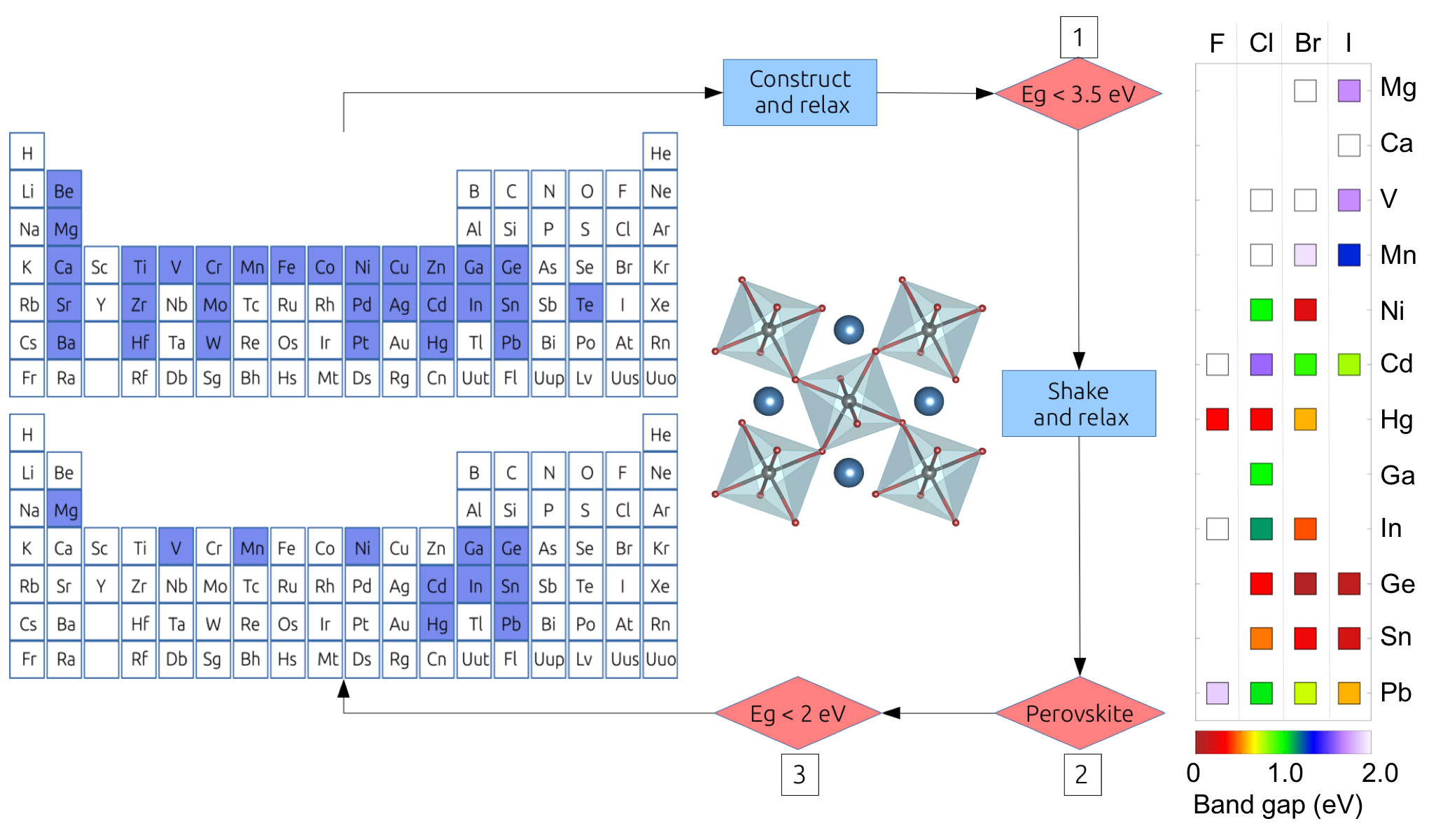}
\caption{{\bf Homovalent replacement of lead.}
Schematic illustration of the methodology and the selection criteria (in red triangles)
employed for screening the Periodic Table for stable lead-free ABX$_3$ halide perovskites. At
first, all perovskite structures with a DFT/LDA band gap larger than 3.5~eV were
eliminated. The remaining compounds were randomly distorted (`shake-and-relax') to probe their
dynamical stability and only compounds which retained the perovskite structure and exhibited a band gap of
less than 2.0~eV  were considered for the final $GW$ calculations. These compounds are marked with the
blue squares in the lower Periodic Table. The DFT/LDA band gaps for all the halide perovskites
which satisfied all selection criteria are summarised at the left side of the figure. Adapted with
permission from \cite{Filip2016}, Copyright (2016) American Chemical Society.
}
\label{fig:1}
\end{center}
\end{figure*}

A comprehensive search for other +2 B-site cations was performed by \cite{Filip2016}.
In this work the Periodic Table was screened by means of DFT calculations, using the procedure
illustrated in FIG.~\ref{fig:1}. The candidate cations include metals which exhibit a stable $+2$ oxidation
state and which form BX$_2$ salts with X = Cl, Br, or I. This choice leads to 116 hypothetical
compounds for which the electronic structure was calculated using DFT/LDA, using the same orthorhombic
perovskite structure as for CH$_3$NH$_3$PbI$_3$. The calculated band structures were then used
to reduce the materials space, and only those insulating compounds with a band gap smaller than
3.5~eV were retained, leaving 40 potential candidates. These compounds were tested for dynamical stability
using a `shake-and-relax' approach, meaning that the structures are randomly distorted and
subsequently re-optimised. Only structures that preserve the perovskite connectivity are
retained in this step. For the 32 compounds that pass this test, more refined calculations
including spin-orbit coupling effects were performed, and 25 hypothetical perovskites with
band gaps smaller than 2~eV were identified. For the compounds with the most promising electronic properties
\cite{Filip2016} performed additional calculations at the $GW$ level. However in all the cases
considered the quasiparticle band gaps were too wide for solar cell applications. This result
suggests that homovalent Pb replacement may not be the best option for eliminating Pb in halide
perovskites. A possible promising exception is the use of Sn$^{2+}$, provided the stability
issues can be resolved. Similar conclusions were reached by an extensive,
high-throughput search of novel ABX$_3$ perovskites by \cite{Korbel2016}.

    \end{subsection}

\begin{subsection}{Heterovalent Pb replacement}

In the previous section we considered the possibility of replacing Pb with another divalent cation.
In order to broaden the search space several authors considered the potential replacement of Pb$^{2+}$
by cations in their +3 or +4 formal oxidation states. This choice leads to the formation of perovskite-related
structures such as for example Cs$_3$Sb(III)$_2$I$_9$ and Cs$_3$Bi$_2$I$_9$~\citep{Saparov2015,Park2015},
Cs$_2$Sn(IV)I$_6$, and Cs$_2$PdBr$_6$~\citep{Lee2014, Sakai2017}. In all these structures the
three-dimensional corner-sharing connectivity of the octahedra is disrupted. In particular, A$_2$BX$_6$
compounds can be described as vacancy-ordered double perovskites, where the B-sites of the perovskite
lattice are replaced alternatively by the +4 cation and by a vacancy. In this structure the BX$_6$ octahedra
are structurally disconnected. Similarly A$_3$B$_2$X$_9$ compounds can be thought of as if obtained
from the perovskite lattice by removing one layer of octahedra. In both cases the resulting compounds
behave electronically as low-dimensional systems (zero-dimensional and two-dimensional, respectively),
and as a consequence the optoelectronic properties are not ideal~\citep{Xiao2017b}.

Another potential strategy
for lead replacement is the split-cation approach. This consists of replacing pairs of Pb$^{2+}$
cations by pairs of +1 and +3 cations, so as to preserve the average oxidation state on the B-site.
When the +1 and +3 cations are arranged in a rock-salt sublattice the structure is an ordered double
perovskite, also known as elpasolite, as shown in FIG.~\ref{fig:2}. In this case the three-dimensional
connectivity of the octahedra is maintained.

Double perovskites are well known in the literature on oxide perovskites~\citep{Vasala2015}.
Halide double perovskites are less known, but are commonly employed in radiation detectors as
scintillators~\citep{Loef2002,Biswas2012}. Common elpasolites exhibit wide band gaps, therefore they
are unsuitable for photovoltaics applications. In the following three sections we review
recent findings on three new classes of halide elpasolites: halide double perovskites
based on pnictogens and noble metals~\citep{Volonakis2016,Filip2016b,
Slavney2016,McClure2016}; indium-silver halide double perovskites~\citep{Volonakis2017, Zhao2017};
and indium-bismuth double perovskites~\citep{Volonakis2017b,Xiao2017,Zhao2017b}.

\begin{figure*}[t!]
\begin{center}
\includegraphics[width=0.7\textwidth]{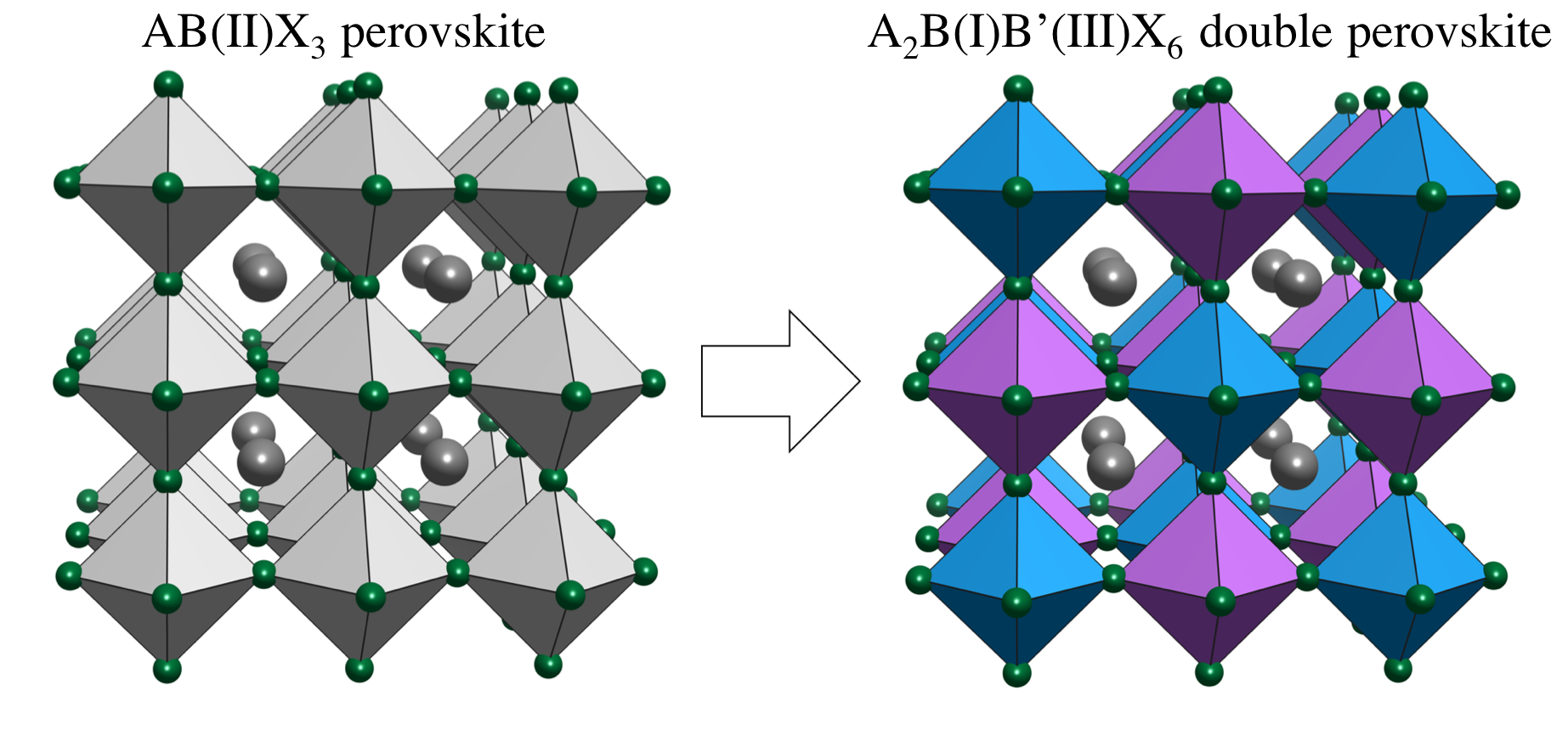}
\caption{{\bf The double perovskite structure.} Two divalent cations at
the B-site of the AB(II)X$_3$ perovskite structure (left), are replaced with a
monovalent and a trivalent cation to form the double perovskite A$_2$B(I)B$^\prime$(III)X$_6$
structure (right). The two-types of octahedra in the double perovskite remain
corner-sharing. When the B(I) and B$^\prime$(III) cations are ordered in a rock-salt sublattice
these compounds are called elpasolites~\citep{Morss1970,Meyer1978,Meyer1980}.}
\label{fig:2}
\end{center}
\end{figure*}

    \begin{subsubsection}{Double perovskites based on pnictogens and noble metals}

\begin{figure*}[t!]
\begin{center}
\includegraphics[width=0.85\textwidth]{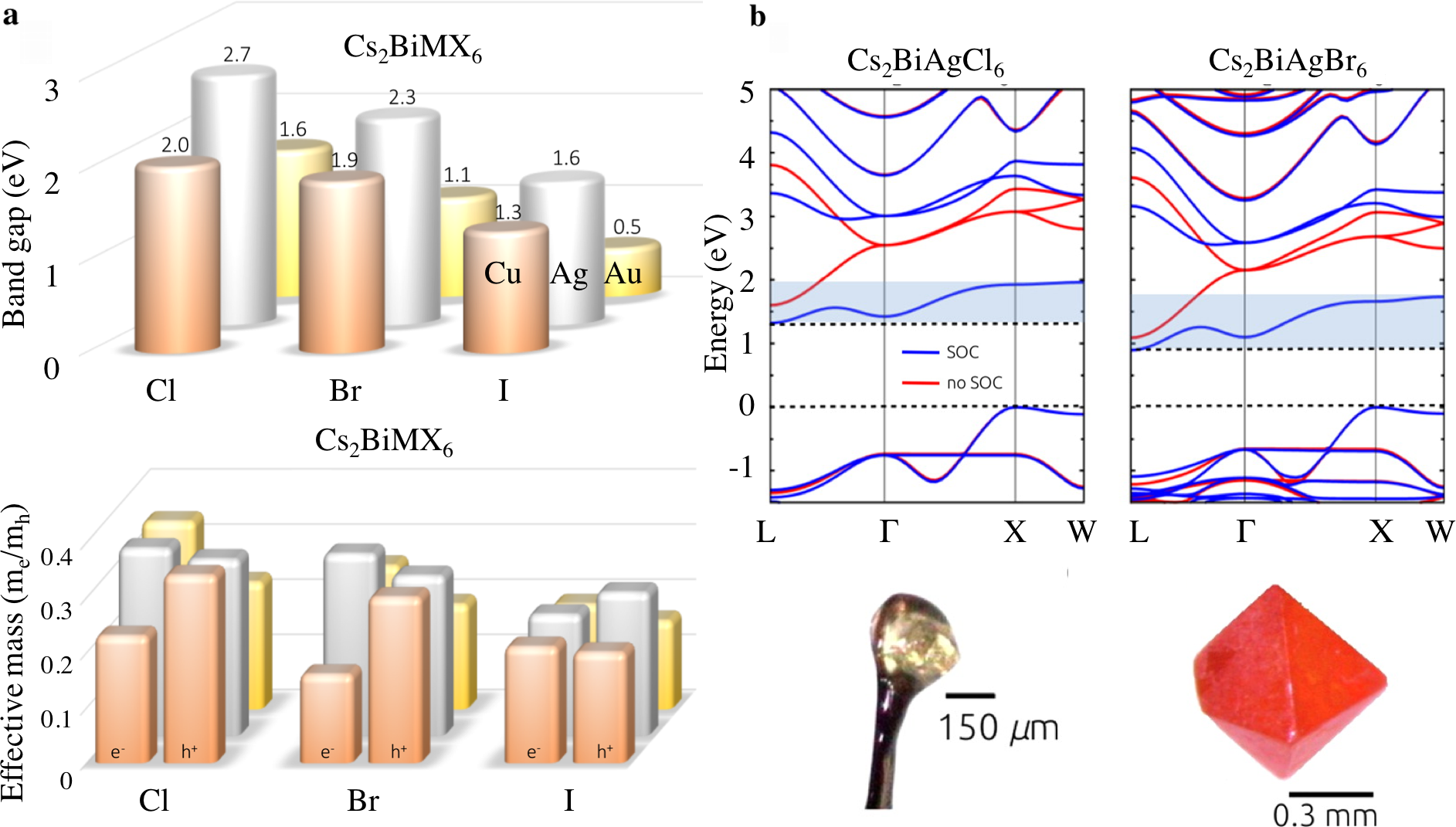}
\caption{{\bf Electronic properties of the pnictogen/noble metal double perovskites.}
{\bf a}. Calculated electronic band gaps (top left) and effective masses (bottom left) of all
the hypothetical Cs$_2$BiB$^\prime$X$_6$ lead-free halide double perovskites,
with B$^\prime$ = Cu, Ag, Au  and X = Cl, Br, I. All the band gaps are calculated by employing the PBE0
hybrid functional, and are found to be indirect and below 2.7~eV. The effective masses are low
($<$0.4~m$_e$), as  calculated  from the DFT/LDA at the valence band top and the conduction band bottom
for the holes and electrons, respectively. 
{\bf b}. The DFT/LDA electronic band structure for the
synthesized Cs$_2$AgBiCl$_6$ and
Cs$_2$AgBiBr$_6$ compounds. Calculations with (without) spin-orbit coupling are shown in
blue (red), and the shaded area highlights the lowest unoccupied band. The synthesized
Cs$_2$AgBiCl$_6$ and Cs$_2$AgBiBr$_6$ single-crystals are shown below. Adapted with
permission from~\cite{Volonakis2016} and \cite{Filip2016b}, Copyright (2016) American Chemical
Society.  
}
\label{fig:3}
\end{center}
\end{figure*}

In order to design new lead-free halide double perovskites \cite{Volonakis2016} started from
the known elpasolite Cs$_2$NaBiCl$_6$~\citep{Morss1972}, and considered the replacement of monovalent Na
by a noble metal. The rationale behind this strategy is that noble metals are known
to be good electrical conductors due to their $s^1$ electrons. \cite{Volonakis2016} performed
DFT calculations for hypothetical compounds Cs$_2$NmPnX$_6$, where Nm = Cu, Ag, Au, Pn = Bi, Sb,
and X = Cl, Br, I. The calculated band gaps and carrier effective masses for the Bi-based
hypothetical perovskites are shown in FIG.~\ref{fig:3}a. All band gaps fall below 2.7~eV, and
the effective masses do not exceed 0.4~$m_e$. These results indicate that pnictogen/noble metal
halide double perovskites might indeed be promising for optoelectronic applications.
\cite{Volonakis2016} reported the synthesis of Cs$_2$AgBiCl$_6$. The compound was synthesized by
following the same route as for Cs$_2$NaBiCl$_6$~\citep{Morss1972}, and replacing AgCl for NaCl.
The structure of this compound was characterized by powder and single crystal X-ray diffraction (XRD)
measurements. The authors obtained an ordered double perovskite structure with the $Fm\overline{3}m$
face centered cubic space-group (no. 255) at room temperature. The reported lattice constant is 10.78~\AA,
in good agreement with the predicted DFT/LDA value of 10.50~\AA. Following the same synthetic route
\cite{Filip2016b} reported  the new double perovskites Cs$_2$AgBiBr$_6$. Also in this case
the XRD measurements confirmed the formation of an elpasolite structure, with a lattice constant
of 11.26~\AA. This value is slightly larger than for Cs$_2$AgBiCl$_6$, as expected based on the
ionic radii of the halides. The synthesis of these compounds via solution and via solid-state synthesis
was also reported independently (and a few weeks earlier) by two other groups \citep{Slavney2016,McClure2016}.

The electronic band structures of the new double perovskites are shown in FIG.~\ref{fig:3}b.
In both cases the  band gap is indirect, with the valence band top at the $X$ point of the
Brillouin zone, and the conduction band bottom at the $L$ point. The band gaps calculated using
the PBE0 hybrid functional are 2.7~eV and 2.3~eV for Cs$_2$AgBiCl$_6$ and Cs$_2$AgBiBr$_6$, respectively
\citep{Volonakis2016}. Quasiparticle $GW$ calculations yield slightly smaller gaps, namely
2.4~eV and 1.8~eV, respectively~\citep{Filip2016b}. The calculated band gaps are broadly in
agreement with the measured optical gaps, which range between 2.2-2.8~eV and 1.9-2.2~eV for
Cs$_2$AgBiCl$_6$ and Cs$_2$AgBiBr$_6$, respectively~\citep{Filip2016b}.

More recently \cite{Filip2017} performed a systematic DFT investigation of the thermodynamic stability
of the entire family of Cs$_2$NmPnX$_6$ compounds. The calculations involve the comparison between
the total energy of each perovskite and the total energies  of all possible decomposition products.
This study predicted that only three double perovskites should be stable, namely
Cs$_2$AgBiCl$_6$, Cs$_2$AgBiBr$_6$, Cs$_2$AgSbCl$_6$, in line with experimental
observations~\citep{Slavney2016,McClure2016,Volonakis2016,Tran2017}.

Among these new compounds, the double perovskite Cs$_2$AgBiBr$_6$ attracted considerable interest
as a potential new material for photovoltaics and optoelectronic applications. In fact, solar
cells using Cs$_2$AgBiBr$_6$ as the active layer were demonstrated by \cite{Greul2017}, and
power conversion efficiencies of 2.5\% were reported. Furthermore, the possibility of mixing
Cs$_2$AgBiBr$_6$ with CH$_3$NH$_3$PbI$_3$ in order to reduce the Pb content of perovskite solar
cells (as opposed to completely replace Pb) is also under investigation~\citep{Du2017b,Du2017}. Finally, \cite{Pan2017} showed
that single crystals of Cs$_2$AgBiBr$_6$ can be employed to realize X-ray detectors with low detection
thresholds~\citep{Pan2017}.

\end{subsubsection}

\begin{subsubsection}{Double perovskites based on In(III) and Ag(I)}

\begin{figure*}
\begin{center}
\includegraphics[width=0.5\textwidth]{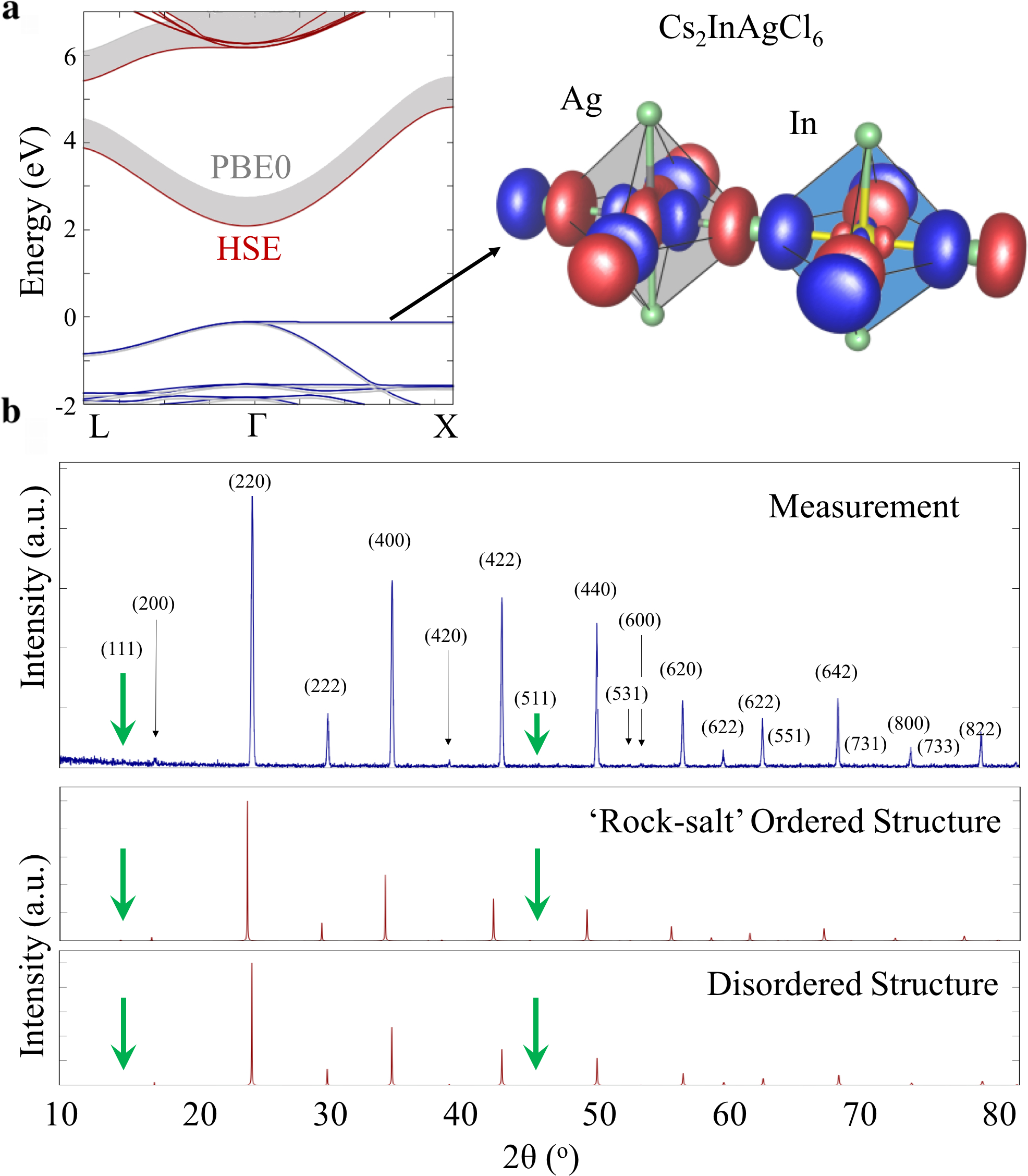}
\caption{{\bf Electronic and structural properties of Cs$_2$InAgCl$_6$.} 
{\bf a}. Electronic
band structure of Cs$_2$InAgCl$_6$ within the DFT/PBE0 and the DFT/HSE hybrid
functionals. Square moduli of the electronic wavefunctions for the highest (top) and second highest
(bottom) occupied state along the $\Gamma$-X direction. 
{\bf b}. Measured powder XRD pattern
of Cs$_2$InAgCl$_6$ (top), XRD pattern calculated by assuming ideal rock-salt ordering of the In
and Ag cations (middle), and XRD pattern calculated by assuming a fully disordered cation
sublattice (bottom). The green arrows show the fingerprints of the cation ordering.
The difference between the calculated ordered and disordered patterns are within the noise level
of the measured spectrum, therefore it is not possible to unambiguously confirm cation ordering
in this case. Adapted with permission from~\cite{Volonakis2017}, Copyright (2017) American Chemical
Society.
}
\label{fig:4}
\end{center}
\end{figure*}

Following up on the successful design and synthesis of lead-free halide double perovskites based on
pnictogen/noble metals combinations, several groups set to improve the design by targeting
double perovskites with a direct band gap. \cite{Volonakis2017} reasoned that the origin of the
indirect band gap of Cs$_2$AgBiBr$_6$ lies in the hybridization between Bi $s$-states and Ag $d$-states
near the valence band top. In order to demonstrate this effect, the authors artificially lowered
the energy of the Bi $s$-states in Cs$_2$AgBiBr$_6$ using an effective Hubbard potential in the
DFT calculations. This strategy led to the appearance of a direct band gap at the zone center,
in line with expectations.
In order to exploit this finding, \cite{Volonakis2017} proposed to replace Bi with a +3 cation
with the outermost $s$-shell unoccupied. This proposal, combined with the observation that the synthetic route
to prepare Cs$_2$InNaCl$_6$ via solution was already known~\citep{Morss1972,Meyer1978},
motivated the consideration of hypothetical compounds of the type Cs$_2$InAgX$_6$ with X = Cl or Br
\citep{Volonakis2017}.

The calculated band structure of Cs$_2$InAgCl$_6$ is shown in FIG.~\ref{fig:4}a. As
expected the system exhibits a direct gap at the $\Gamma$ point, and the top of the valence band
is comprised of Ag and In $d$-orbitals, hybridized with the halogen $p$-orbitals.
The band gap of Cs$_2$InAgCl$_6$ was predicted to be in the range 2.7$\pm$0.6~eV, with
the error bar reflecting the spread in band gaps obtained from calculations based on
different exchange and correlation functionals. This compound was successfully synthesized via
an acidic solution synthesis route by mixing stoichiometric amounts of InCl$_3$, AgCl and CsCl.
Powder XRD measurements confirmed the formation of a double perovskite with a face-centered cubic
crystal ($Fm\overline{3}m$ space group) and a lattice constant of 10.47~\AA, in good agreement
with the DFT predictions~\citep{Volonakis2017}.

The nature of In/Ag cation ordering in Cs$_2$InAgCl$_6$ is not firmly established yet.
The characteristic fingerprints of cation ordering in the XRD spectra of double perovskites
are two doublets arising from reflections at the planes containing each cation (FIG.~\ref{fig:4}b).
In contrast to the cases of Cs$_2$AgBiCl$_6$ and Cs$_2$AgBiBr$_6$, where such doublets have clearly
been identified \citep{Volonakis2016}, the assignment proved more difficult for Cs$_2$InAgCl$_6$.
\cite{Volonakis2017} calculated the XRD patterns for structures with ordered or disordered cations
and compared these results to measured spectra, as shown in FIG.~\ref{fig:4}b. From this
comparison it is clear that the difference between the two structures is too subtle
to be resolved by the measurements. More refined XRD data from~\cite{Zhou2017},
taken on single crystal samples, show a clear doublet at low angle. This supports
the notion that Cs$_2$InAgCl$_6$ is a fully-ordered elpasolite, although further investigations
would be desirable to settle this question.

The calculated electron and hole effective masses of Cs$_2$InAgCl$_6$ are relatively light,
0.29~$m_e$ and 0.28~$m_e$, respectively. These values refer to the parabolic bands that can
be seen in FIG.~\ref{fig:4}a. The non-dispersive band that is seen along the $\Gamma X$ direction
in the same figure arises from two-dimensional wavefunctions comprising of
In 4d$_{x^2-y^2}$ states and Cl 3$p_{x,y}$-states. These states are expected to hinder hole transport
along the six equivalent [001] directions, as well as to favor the formation of deep traps.

The nature of the band gap of Cs$_2$InAgCl$_6$ remains an open question. The absorption data
reported by \cite{Volonakis2017} indicate an absorption onset around 3.3~eV,
which is consistent with the upper range obtained from DFT calculations.
However the photoluminescence data by \cite{Volonakis2017} also indicate emission near 2.0~eV,
indicating the presence of optically-active defects within the gap.
\cite{Jiajun2017} proposed an alternative explanation for these
effects, involving parity-forbidden band-to-band transitions.

While the band gap of Cs$_2$InAgCl$_6$ is too wide for applications in photovoltaics, this
new double perovskites is attracting attention for applications in UV detectors
due to its excellent stability and non-toxicity.
In particular \cite{Jiajun2017} and \cite{Zhou2017} succeeded
in growing mm-sized single crystals, and \cite{Jiajun2017}
fabricated UV detectors with high on/off ratios, fast photoresponse, low dark current, and
high sensitivity.

\end{subsubsection}

\begin{subsubsection}{Double perovskites based on Bi(III) and In(I)}

\begin{figure*}[t!]
\begin{center}
\includegraphics[width=0.8\textwidth]{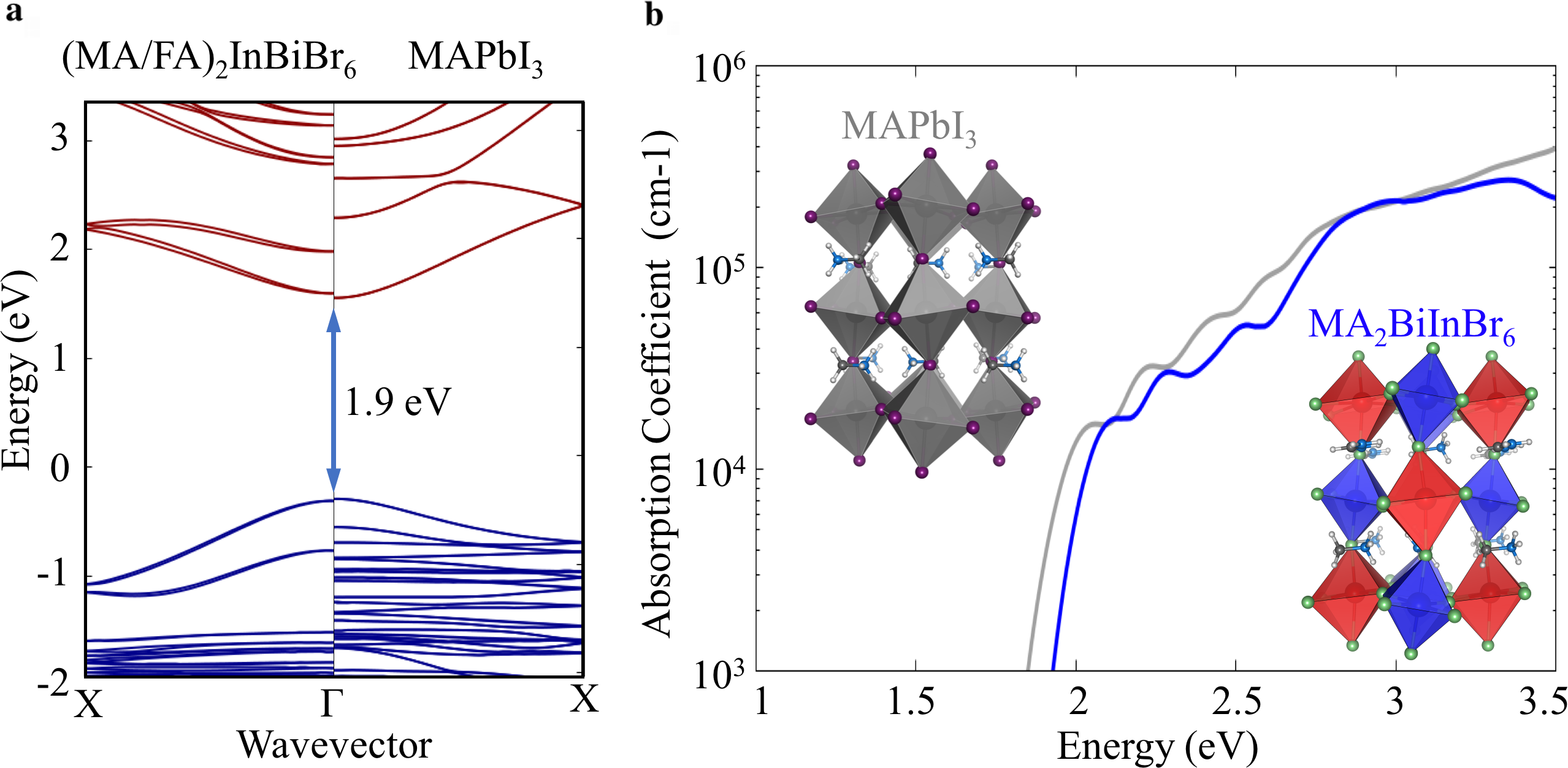}
\caption{{\bf Indium/Bismuth halide double perovskites.} 
{\bf a}. Comparison between the electronic
band structures of the hypothetical lead-free double perovskite (MA/FA)$_2$InBiBr$_6$ and the lead
halide perovskite MAPbI$_3$. MA stands for methylammonium (CH$_3$NH$_3$), and FA for formamidinium
[CH(NH$_2$)$_2$]. Calculations were performed by employing the PBE0 hybrid
functional. The double perovskite structures were fully optimized starting from the
low-temperature $Pnma$ phase of MAPbI$_3$, and by replacing two Pb atoms with In and Bi.
{\bf b}. Calculated absorption coefficients for MA$_2$InBiBr$_6$ (blue line), and MAPbI$_3$ (grey line). 
Adapted with permission from~\cite{Volonakis2017b}, Copyright (2017) American Chemical Society.
}
\label{fig:6}
\end{center}
\end{figure*}

The new double perovskites described in the preceding sections either exhibit indirect band gaps,
or non-dispersive bands at the band extrema. In order to move a step closer to realizing
high-quality semiconducting halide perovskites without lead, \cite{Volonakis2017b} revisited
the general design principles of the split-cation approach based on +1 and +3 cations.

The authors argued that an ideal double perovskite should contain cations with the same valence
configuration as Pb$^{2+}$. This implies that candidate elements should have
(i) occupied $s$-orbitals, (ii) unoccupied $p$-orbitals, and (iii) a filled
$d$ shell far from the band edge.
The candidates for the monovalent cation are limited to the alkali metals,
the noble metals and the Group~III elements (boron group). Among these options only the
latter have occupied $s$-orbitals and unoccupied $p$-orbitals in their formal +1 oxidation state.
However, only two members of this group, In and Tl, are stable as +1 cations, and Tl is to be
excluded due to its toxicity. This leaves In$^+$ as the sole candidate for the monovalent B-site
cation.
For the trivalent B-site cation the choice is between pnictogens, some transition metals such
as  Sc, Y, Cu and Au that were previously reported to form elpasolites~\citep{Giustino2016},
elements of the boron group in the +3 oxidation state, lanthanides and actinides. Among these
elements only the pnictogens satisfy the criteria (i)-(iii) above. In this group N and P are
too small to coordinate six halogens in a octahedral arrangement, and As is toxic, therefore
the only options left are Sb$^{3+}$ and Bi$^{3+}$.
Based on this simple reasoning \cite{Volonakis2017b} concluded that the best option for
replacing Pb, from the electronic structure viewpoint, is to consider In(I)/Sb(III) and
In(I)/Bi(III) double perovskites.

Building on this insight, \cite{Volonakis2017b} evaluated the stability of the hypothetical compounds
A$_2$InPnX$_6$ with A = K, Rb, Cs, Pn = Sb, Bi, and X = F, Cl, Br, I, using both
the Goldschmidt tolerance analysis and DFT calculations of decomposition energies in
conjunction with the Materials Project Database~\citep{materialsproject}.
The outcome of this study was that all compounds are
unstable with respect to decomposition, in line with the work of ~\cite{Xiao2017}.
However Cs$_2$InBiBr$_6$ was found to be only marginally unstable,
as the calculated decomposition energy of 1~meV per formula unit falls within the uncertainty
of the computational method. This finding raises the hope that Cs$_2$InBiBr$_6$ or a closely related compound
might be amenable to synthesis. \cite{Volonakis2017b} also noted that the decomposition
energy correlates with the size of the A-site cation, and that double perovskites with
larger A-site cations tend to be less prone to decomposition. Based on this observation the
authors proposed that In(I)/Bi(III) halide double perovskites might be amenable for synthesis by using
large organic cations such as CH$_3$NH$_3^+$ and CH(NH$_2$)$_2^+$.

The electronic band structure of the hypothetical (CH$_3$NH$_3$/CH(NH$_2$)$_2$)$_2$InBiBr$_6$
double perovskites along with the band structure of CH$_3$NH$_3$PbI$_3$ are shown in FIG.~\ref{fig:6}a. 
As anticipated, owing to the similar
electronic configuration of Pb$^{2+}$, In$^+$ and Bi$^{3+}$ the two compounds exhibit
very similar band structures.
The band gaps are direct in both cases and of the same magnitude. Similarly
the calculated optical absorption spectra for these compounds are very similar, as shown in
FIG.~\ref{fig:6}b. These findings highlight the potential of hybrid In/Bi halide double
perovskites for photovoltaics and optoelectronics.

At the time of writing this chapter synthetic attempts at making In(I)/Bi(III) halide double perovskites
have not been successful. The key challenge appears to be the tendency of In(I) to be oxidized to
the more stable form In(III), therefore future synthesis studies should devise strategies for maintaining
In in its +1 state long enough to enable its incorporation in the double perovskite lattice.
\end{subsubsection}

\end{subsection}

\end{section}

\begin{section}{Conclusions}

In this chapter we have presented a detailed account of the contributions made by our materials
modelling group to the study of the fundamental properties and design principles of metal-halide
perovskites. The results reviewed in this chapter were obtained by employing techniques ranging from
standard density functional theory and density functional perturbation theory to many-body
perturbation theory treatments of electronic excitations and electron-phonon coupling phenomena. Our 
studies so far have been directed at two principal goals, understanding the fundamental properties 
of lead-halide perovskites and developing design principles to guide the discovery of novel 
lead-free metal-halide perovskite semiconductors.

The current understanding of the fundamental properties of lead-halide perovskites is due to a
complementary effort combining state-of-the art predictive computational modelling and experimental
measurements which have allowed us to rationalize the optoelectronic properties of these materials
from the perspective of atomistic modelling. In addition, we have shown that the predictive power of
computational modelling can be directed at designing new lead-free halide perovskites, which led
most notably to the computational discovery of a new family of lead-free halide double
perovskites. During the last year more than eight
novel lead-free double perovskite compounds have been successfully synthesized
\citep{Volonakis2016,Slavney2016,McClure2016,Filip2016b,Volonakis2017, Tran2017,Wei2016,Deng2016,
Wei2017}. Among these, compounds such as Cs$_2$AgBiCl$_6$, Cs$_2$AgBiBr$_6$, Cs$_2$InAgCl$_6$ and
Cs$_2$AgSbCl$_6$, were first proposed by us {\it in silico}. 

Despite the tremendous popularity of halide perovskites, there are still
fundamental open questions and technological challenges that remain unresolved. A non-exhaustive 
list includes the fundamental understanding of charge transport and excitonic properties, 
elucidating the physics of defects and their impact on the degradation of these materials 
under standard solar cell operating conditions, and developing a practical solution for  
reducing the Pb content  without impacting the device performance. All of these problems are 
currently active areas for both experimental and theoretical research.

In a broader context, the rapid rise of hybrid halide perovskites can be seen as a true success 
story for experimental and theoretical materials design. As the race for ever higher performing and 
cost-efficient optoelectronic devices continues, there is an increasing demand for the discovery  of 
new functional materials. Thanks to the continued development of highly accurate {\it ab initio} 
methods, computational materials modelling can now directly guide  discovery, by designing materials 
and predicting their properties {\it in silico} ahead of synthesis. These developments 
place {\it ab initio} modelling in a central role in modern materials science and engineering, as an 
accelerator for the discovery and deployment of new technologies.

\end{section}

\begin{section}{Acknowledgement}

The research leading to these results has received
funding from the Graphene Flagship (Horizon 2020 Grant
No. 696656 - GrapheneCore1), the Leverhulme Trust (Grant RL-2012-
001), and the UK Engineering and Physical Sciences Research
Council (Grant No. EP/J009857/1, EP/M020517/1 and EP/
L024667/1).
\end{section}

%
%
\end{document}